\PassOptionsToPackage{dvipsnames,svgnames}{xcolor}

\documentclass[screen,sigconf]{acmart}

\copyrightyear{\copyright{} Zeyu Huang, Zuyu Xu, Yuanhao Zhang, Chengzhong Liu, Yanwei Zhao, Chuhan Shi, Jason Chen Zhao, Xiaojuan Ma, 2025. This is the author's version of the work. It is posted here for your personal use. Not for redistribution. The definitive Version of Record is published in ACM Digital Library.}
\acmYear{2025}
\setcopyright{none}
\acmConference[ICHEC 2025]{Proceedings of the 2025 International Conference on Human-Engaged Computing}{November 21--23, 2025}{Singapore, Singapore}
\acmBooktitle{Proceedings of the 2025 International Conference on Human-Engaged Computing (ICHEC 2025), November 21--23, 2025, Singapore, Singapore}
\acmPrice{}
\acmDOI{10.1145/3786995.3786999}
\acmISBN{979-8-4007-2234-9/2025/11}

\acmSubmissionID{11}

\usepackage{pifont}
\usepackage{longtable}
\usepackage{fixfoot}
\usepackage{mwe}
\usepackage{makecell}
\usepackage[inline]{enumitem}
\usepackage{xspace}
\usepackage{multirow} 
\usepackage{caption}
\usepackage{subcaption}
\usepackage{tabularx} 
\usepackage{booktabs} 
\usepackage{CJKutf8}
\usepackage{hyperref}
\usepackage[nameinlink,noabbrev]{cleveref}

\newcommand{\edit}[1]{{#1}}

\newcommand{\company}{\anon{FieldFusion Ltd.}}
\newcommand{\hkust}{%
  \affiliation{%
    \institution{The Hong Kong University of Science and Technology}%
    \city{New Territories}
    \state{Hong Kong}%
    \country{China}%
  }%
}
\newcommand{\atconnect}{@connect.ust.hk}

\begin{document}

\title[Mixed Reality Scenic Live Streaming for Cultural Heritage]{Mixed Reality Scenic Live Streaming for Cultural Heritage: Visual Interactions in a Historic Landscape}


\author{Zeyu Huang}
\email{zhuangbi\atconnect}
\orcid{0000-0001-8199-071X}
\hkust

\author{Zuyu Xu}
\email{zxubo\atconnect}
\orcid{0009-0001-5138-9285}
\hkust

\author{Yuanhao Zhang}
\email{yzhangiy\atconnect}
\orcid{0000-0001-8263-1823}
\hkust

\author{Chengzhong Liu}
\email{chengzhong.liu\atconnect}
\orcid{0000-0002-4948-8276}
\hkust

\author{Yanwei Zhao}
\email{458815330@qq.com}
\orcid{0009-0000-2694-581X}
\affiliation{%
  \institution{Zhejiang University}%
  \city{Hangzhou}%
  \state{Zhejiang}%
  \country{China}%
}

\author{Chuhan Shi}
\email{chuhanshi@seu.edu.cn}
\orcid{0000-0002-3370-1626}
\affiliation{%
  \institution{Southeast University}%
  \city{Nanjing}%
  \state{Jiangsu}%
  \country{China}%
}

\author{Jason Chen Zhao}
\email{zhaoch21@gmail.com}
\orcid{0000-0002-0840-8022}
\affiliation{%
  \institution{FieldFusion Ltd.}
  \city{Hangzhou}
  \state{Zhejiang}
  \country{China}
}

\author{Xiaojuan Ma}
\authornote{Corresponding author}
\email{mxj@cse.ust.hk}
\orcid{0000-0002-9847-7784}
\hkust







\begin{abstract}
  Scenic Live Streams (SLS), capturing real-world scenic sites from fixed cameras without streamers, have gained increasing popularity recently. They afford unique real-time lenses into remote sites for viewers' synchronous and collective engagement. Foregrounding its lack of dynamism and interactivity, we aim to maximize the potential of SLS by making it interactive. Namely MRSLS, we overlaid plain SLS with interactive Mixed Reality content that matches the site's geographical structures and local cultural backgrounds. We further highlight the substantial benefit of MRSLS to cultural heritage site interactions, and we demonstrate this design proposal with an MRSLS prototype at a UNESCO-listed heritage site in China. The design process includes an interview (N=6) to pinpoint local scenery and culture, as well as two iterative design studies (N=15, 14). A mixed-methods, between-subjects study (N=43, 37) shows that MRSLS affords immersive scenery appreciation, effective cultural imprints, and vivid shared experience. With its balance between cultural, participatory, and authentic attributes, we appeal for more HCI attention to (MR)SLS as an under-explored design space.
\end{abstract}






\maketitle

\section{Introduction}

The ways we experience cultural heritage sites are rapidly evolving beyond physical presence. Digital mediation, from online archives~\cite{koukopoulosEvaluatingUsabilityPersonal2019,vertUserEvaluationMultiPlatform2021,kotutTrailHeritageSafeguarding2021,burkeyTotalRecallHow2019} to virtual tours~\cite{langConstructionIntangibleCultural2019,romanVirtualSpaceTourism2022,tungAugmentedRealityMobile2015,siddiquiVirtualTourismDigital2022}, offers valuable windows into distant landscapes, a trend accelerated by the recent global pandemic.
Yet, a tension persists within this digital landscape. On one hand, immersive Extended Reality (XR) applications provide lifelike, yet often solitary, encounters~\cite{langConstructionIntangibleCultural2019,romanVirtualSpaceTourism2022,tungAugmentedRealityMobile2015,siddiquiVirtualTourismDigital2022}. On the other, community-driven platforms foster a sense of collective presence but can lack deep, engaging interaction with the site itself~\cite{koukopoulosEvaluatingUsabilityPersonal2019,vertUserEvaluationMultiPlatform2021,kotutTrailHeritageSafeguarding2021,burkeyTotalRecallHow2019}.
This leaves a crucial gap: an experience that combines the liveness of being there and the richness of communal engagement, creating a more complete ``virtual trip'' experience.

In this paper, we turn our attention to the recent popularity of scenic live streams (SLS), sometimes also known as \emph{world cameras} or \emph{live webcams}. They broadcast real-world landscapes in real time through fixed cameras and operate 24/7 non-stop without human streamers.
They can be found on both dedicated websites like SkylineWebcams\footnote{\url{https://www.skylinewebcams.com/}} and general-purpose video platforms like YouTube\footnote{An example Venice SLS: \url{https://www.youtube.com/watch?v=HpZAez2oYsA}}.
With their real-time fidelity and the inherent collective nature of live streaming~\cite{huWhyAudiencesChoose2017, wulfWatchingPlayersExploration2020,huangBeThereBe2024}, SLS have been increasingly adopted by cultural heritage sites for their touristic and educational potential~\cite{jarrattExplorationWebcamtravelConnecting2021,jarrattWebcamtravelConceptualFoundations2021}.
However, the conventional SLS experience remains a passive one, largely confined to observational appreciation.
This passivity not only fails to communicate the layered cultural narratives and living heritage embedded within these sites~\cite{jarrattExplorationWebcamtravelConnecting2021}, but also misses the opportunity to leverage the participatory, community-building power of the live stream medium~\cite{hamiltonStreamingTwitchFostering2014}.

To unlock the full potential of SLS for remote cultural heritage engagement, we introduce \emph{Mixed Reality Scenic Live Streams} (MRSLS), a new form of interactive multimedia experiences derived from regular SLS that dynamically overlays interactive virtual elements onto the live scenic footage.
By semantically aligning these digital augmentations with the site's geographical structures and cultural context, MRSLS elevates passive viewing into an active, participatory, and culturally resonant experience.
This fusion of the real and the virtual forges new pathways for cultural transmission and digital placemaking, connecting global audiences in real time with the symbols, traditions, and stories of a place.

We demonstrate this concept through the design and evaluation of an MRSLS prototype at a renowned UNESCO-listed heritage landscape in China~\cite{xuSignificanceWestLake2017, zhangCulturalLandscapeMeanings2019,worldheritagecentreStateConservationProperties2019}.
Our generalizable design workflow, which involved interviews to formalize local elements and iterative feature refinement, produced a prototype with context-aware visual overlays and interactive features tailored to the site's unique picturesque and literary charms.
A subsequent mixed-methods, between-subjects study (N=43, 37) confirmed that MRSLS effectively affords a sense of virtual embodiment, deepens cultural appreciation, and fosters social connection among remote viewers.

In sum, this research makes three primary contributions:
\begin{enumerate*}
\item We introduce and formalize the concept of MRSLS as a new interaction mechanism for cultural heritage sites.
\item We provide an exemplary design flow and a functional MRSLS prototype for a real-world heritage site.
\item We derive design implications and future research directions for the application and generalization of MRSLS.
\end{enumerate*}

\section{Related Work}

\subsection{XR for Cultural Heritage Sites}


Virtual Reality (VR) is a dominant approach for creating interactive, remote experiences of cultural heritage sites. Typically, this involves meticulously reconstructing a site in a virtual environment to allow for free exploration~\cite{siddiquiVirtualTourismDigital2022,romanVirtualSpaceTourism2022,scharffhausenWhatIfWe2024,tongApplyingCinematicVirtual2024,gattoImprovingAccessibilityCultural2025}.
Some systems enhance this by incorporating multi-user features, fostering lively social interactions within these virtual replicas~\cite{iacovielloHoloCitiesSharedReality2020,luRevivingEustonArch2023}.
While immersive, VR reconstructions are inherently static. They capture a single moment in time, often missing a site's authentic, real-time dynamism---the daily rhythm of local life, the serendipitous cultural moments, etc.
Recreating this ``livingness'' in VR requires immense effort in 3D animation and continuous updates to reflect the present.

\edit{
  Beyond reconstruction-centric VR, narrative guides and digital storytelling have also been used to deliver curated narratives at heritage sites and museums.
  \citet{sylaiouVirtualHumansMuseums2022} present systems where virtual docents mobilize speech, gaze, and gestures to create attractive heritage storytelling and improve social presence.
  \citet{arendttorpGrabItYou2023} report a co-designed VR narrative for intangible heritage, where indigenous storytellers and visitors engage in embodied mutual interactions to strengthen authenticity and interest.
  Recent heritage chapters also note pilots of live-streamed virtual tours where an avatar shows around the scenes and answers questions in in real time~\cite{hutsonImmersiveTechnologies2024}.
  These storytelling applications focus more on informativeness, and sometimes provide a companion narrator. However, they cannot offer first-person immersion at the broadcast site or direct interactions with it.
}

\edit{In contrast to full immersion in virtual replicas or augmenting online guides with touristic information, what we want to achieve is an engaging, participatory, and situated experience of cultural heritage sites, foregrounding the site's real-time ``livingness''.}
We posit that Mixed Reality (MR) offers a compelling alternative by overlaying interactive digital content onto an authentic, live view of a heritage site.
A primary challenge for \emph{remote} MR is accessing a real-time visual feed of the physical environment.
Inspired by the recent popularity of touristic live streams~\cite{zhangCanLiveStreaming2021,linLiveStreamingTourism2022,qiuCanLiveStreaming2021,dengBlendedTourismExperiencescape2019,luVicariouslyExperiencingIt2019}, we propose that Scenic Live Streams (SLS) can be a promising base for remote MR interactions, for their unmediated, streamer-less, real-time footage of remote sites serves as an authentic visual canvas.
In addition, its typically passive viewership presents a clear design opportunity, a space that remains largely under-explored in HCI~\cite{jarrattExplorationWebcamtravelConnecting2021}.
We envision that enhancing SLS with MR can create more engaging and cultural experiences, analogous to how on-site MR applications already enrich physical visits with historical and cultural cues~\cite{jacobCollaborativeAugmentedReality2021,spierlingExperiencingPresenceHistorical2017,cejkaHybridAugmentedReality2020,wangIntangibleCulturalHeritage2018}.\

Technologically, prior work has demonstrated the feasibility of applying environment-aware MR to fixed-camera footage.
For instance, MR has been overlaid on surveillance cameras to enable interactive monitoring~\cite{szentandrasiPOSTERINCASTInteractive2015} and on machine tools to preview outputs safely~\cite{liuAugmentedRealityassistedIntelligent2017}, using computer vision to make virtual content aware of scene physics.
Building on these precedents, we explore the design of what we term Mixed Reality Scenic Live Streams (MRSLS), aiming to create \edit{engaging remote interactions with cultural heritage sites that emphasize scenery itself as the primary subject and also advocate real-time participation.}

\subsection{Scenic Live Streams as a Design Space}

Scenic Live Streams offer viewers a real-time window into a remote location, fostering a sense of telepresence of ``almost being there'' and ``see through one's eyes''~\cite{baishyaYourEyesAnytime2017,huangBeThereBe2024}.
Consequently, SLS has become a popular tool for cultural heritage sites to maintain touristic and educational outreach, especially when physical travel is constrained~\cite{jarrattExplorationWebcamtravelConnecting2021}.

However, the engagement in most popular live streams hinges not just on liveness, but on its capacity for serendipity and unpredictable dynamics, which conveys a powerful sense of liveness and authenticity~\cite{haimsonWhatMakesLive2017}.
In most live stream genres, this is actively cultivated by human streamers' active mediation---their narration, interaction, and direction of attentio~\cite{luYouWatchYou2018}.
In this sense, traditional SLS by design falls short of its potential.
Its fixed, uncurated perspective, while authentic, can lead to a passive viewing experience, less distinguishable from pre-recorded long scenic videos.
Furthermore, insights from other culture-themed streams, like craftsmanship, suggest that the vibrant atmosphere and lively viewer-viewer interactions in streamer-led live streams is particularly potent in strengthening audience's connection to the cultural cause~\cite{luFeelItMy2019}.

\begin{figure*}[t]
  \centering
  \includegraphics[width=0.95\textwidth]{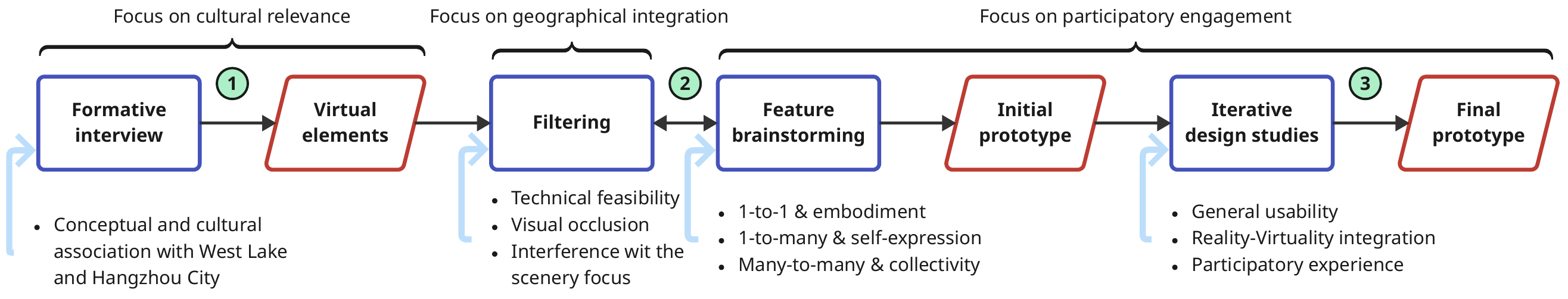}
  \caption{Our design process for an MRSLS at cultural heritage sites.}\label{fig:design-process}
  \Description{A multi-stage design process flow diagram segmented in three parts. The first part focuses on cultural relevance. It contains a formative interview on conceptual and cultural association with West Lake and Hangzhou City, which produces virtual elements. The second part focuses on geographical integration. It contains filtering the virtual elements by technical feasibility, visual occlusion, and interference with the scenery focus. The third part focuses on participatory engagement. It contains feature brainstorming that produces an initial prototype, and iterative design studies that produce a final prototype.}
\end{figure*}

Recognizing both the gaps and potential of SLS, we aim to make it more adept at promoting cultural heritage sites.
We draw particular inspiration from the emergent genre of ``stream chat games'' (e.g., Kukoro\footnote{\url{https://heynaugames.com/kukoro-twitch-interactive-chat-games}}).
These games cleverly enable interactive, audience-led experiences within the boundary of the conventional live stream interface (i.e., via live comments and gifts).
This model of audience's proactive involvement has been shown to boost engagement and social connectedness~\cite{robinsonChatHasNo2022}.
This paper adapts a similar interactive paradigm, investigating how it can be specifically tailored to enrich remote interactions with cultural heritage sites on top of SLS.

\section{Design}\label{sec:design}

To demonstrate the concept of MRSLS as an immersive interactive experience and its application to cultural heritage sites,
we design a concrete MRSLS prototype at West Lake, Hangzhou, China. It is one of the best-known scenic sites across China, and it has been included in UNESCO's World Heritage List in 2011 for its picturesque aesthetic values and profound cultural imprint~\cite{worldheritagecentreStateConservationProperties2019}.
This section elaborates on the iterative design process of this prototype (\cref{fig:design-process}), which balances
\begin{enumerate*}
\item geographical integration into the scene as an MR system,
\item match with the local culture as a cultural heritage experience,
\item participatory engagement as a live stream.
\end{enumerate*}

\subsection{Choosing Interactive MR Content}

Our design started with an interview to formulate a collection of concepts that people tend to relate to West Lake heritage site and the local cultural region (\cref{fig:design-process} \ding{192}).
We recruited \edit{six participants aged 21--23, including four local Hangzhou residents and two non-local Chinese who knew and wanted to visit the place. All six participants are students and have been frequently viewing live streams in their daily lives. We did not particularly screen users with more knowledge about \emph{scenic} live streams or mixed reality, because we wanted to primarily focus on the content and divergent thinking~\cite{cropleyPraiseConvergentThinking2006} at this stage.}

The semi-structured interview mainly covered objects (tangible or intangible, natural or cultural) and activities (physically feasible or not) they tend to associate with West Lake and Hangzhou city.
We deductively coded their inputs by these four categories and their identity (local vs.\@ non-local)~\cite{braunUsingThematicAnalysis2006} in \cref{tab:formative-study}.

\begin{table*}[ht]
  \caption{Formative study participants' reflection on potential elements for an MRSLS at West Lake.}\label{tab:formative-study}
  \footnotesize
  \begin{tabular}{p{2.2cm} p{5cm} p{6cm}}
    \toprule
    & Local People & People Who Want to Travel There \\
    \midrule
    Culture
    & Ancient Chinese poems\newline
    Tea culture\newline
    Oiled-paper umbrella\newline
    History of the Song Dynasty\textsuperscript{c}\newline
    History of the Spring and Autumn Period\textsuperscript{a}
    & Poems from the Tang Dynasty\textsuperscript{b} and Song Dynasty\textsuperscript{c}\newline
    Oiled-paper umbrella\newline
    Legend of the White Snake\textsuperscript{d}\\
    \midrule
    Nature
    & Lotus\newline
    Rain\newline
    Evening
    & Lotus\newline
    Fish\newline
    Rain\newline
    Sunset \\
    \midrule
    Realistic Activities
    & Rowing\newline
    A quiet tour of West Lake
    & Rowing\newline
    Drone aerial photography of the West Lake \\
    \midrule
    Fantasized Activities
    & Watching fireworks\newline
    Looking at West Lake from other angles\newline
    Diving into the West Lake\newline
    Recreating the Legend of the White Snake\textsuperscript{d}\newline
    Virtually going there in an avatar
    & Swimming in the West Lake\newline
    Seeing non-local creatures (polar bears, penguins)\newline
    Virtually going there in an avatar\newline
    Recreating the legend of the White Snake\textsuperscript{d}\\
    \midrule
    \multicolumn{3}{l}{
      \scriptsize
      Note \quad
      \begin{minipage}[t]{0.8\linewidth}
        Spring and Autumn Period\textsuperscript{a}: 770--481 BCE\\
        Tang Dynasty\textsuperscript{b}: 618--907\\
        Song Dynasty\textsuperscript{c}: 960--1279\\
        The Legend of the White Snake\textsuperscript{d}: \url{https://en.wikipedia.org/wiki/Legend_of_the_White_Snake}
      \end{minipage}
    }\\
    \bottomrule
  \end{tabular}
\end{table*}

\edit{These candidate MR elements were subsequently filtered (\cref{fig:design-process} \ding{193}) in two stages.
In the first stage, we detailed the needs and requirements of the elements with regard to the three high-level directions mentioned at the beginning of this section: for geographical integration, certain elements must be relevant to the West Lake surface; for cultural alignment, elements should cater to the two most significant local cultural aspects, literature~\cite{zhangCulturalLandscapeMeanings2019, geSoundscapeWestLake2013} and craftsmanship~\cite{gaoServiceDesignDestination2022, jinExperientialAuthenticityHeritage2020, renRedesignTraditionalChinese2013}; for participatory engagement, certain elements should be relevant to local collective activities.}
In the second stage, we examined the candidates against three more criteria:
\begin{enumerate*}
\item technical feasibility (e.g., those perspective-changing elements were not possible through one fixed camera),
\item visual occlusion of the scenic view (should not be too much),
\item  how much the feature itself diverted the overall theme away from scenery appreciation (e.g., theatrical recreation of legends caused too much diversion), and
\item the visual harmony between the elements and the lake scene.
\end{enumerate*}
The final elements we chose are listed below with their corresponding reasons:

\begin{description}
  \item[Oil-Paper Umbrellas]
    They were a common accessory used by local Hangzhou people in ancient times and have turned into a decorative artifact and a famous form of local cultural art that reflects fine folk craftsmanship~\cite{gaoServiceDesignDestination2022, jinExperientialAuthenticityHeritage2020, renRedesignTraditionalChinese2013}.
    Interviewees in the formative study mentioned that oil-paper umbrellas symbolize romance, reunion, and happiness in Chinese culture~\cite{renRedesignTraditionalChinese2013}.
    Thus, they may assist in facilitating the emotional resonance among viewers.
  \item[Classical Chinese Poetry]
    Classical Chinese poetry is a form of traditional Chinese literature~\cite{xuHowImagesInspire2018}.
    Poems in ancient dynasties have archived many historical and cultural elements of the Hangzhou city, as well as poetic descriptions of and appreciation for the landscape around West Lake~\cite{zhangCulturalLandscapeMeanings2019, geSoundscapeWestLake2013}.
  \item[Lotus]
    The lotus is the representative flower of the West Lake.
    One of the top 10 scenes of West Lake is ``Lotus in the Breeze at Crooked Courtyard''~\cite{geRelationshipSoundscapeChinese2012}.
    Although this flower is not exclusive to the scenic site,
    the abundant Chinese literature that praised lotus on the West Lake contributed to such a cultural connection~\cite{shiCharacteristicsDevelopmentSignificance2015}.
    It is also one of the common shapes of water lanterns Chinese people use.
    Plus, lotus blooms in the summer, right when this research was conducted.
  \item[Fish]
    Similarly, another top scene of West Lake, ``Fish Viewing at the Flower Pond''~\cite{geRelationshipSoundscapeChinese2012}, makes fish a widely recognized ornamental symbol of the West Lake, as confirmed by the interviewees.
  \item[Fireworks]
    Some interviewees shared novel and even imaginative activities that they would like to experience on West Lake.
    Among them, we chose fireworks because they bear fairly universal aesthetic values to ensure the acceptability of a broad audience.
\end{description}

Given all the elements to include, we continued to brainstorm the feature set of the MRSLS prototype (\cref{fig:design-process} \ding{194}), which is detailed in the next subsection.

\subsection{Feature Design and Iteration Protocol}

\subsubsection{Initial Design}\label{sec:initial-design}

When we formulate the feature set, we especially strove to include different viewer-viewer interaction patterns in live streams' participatory experiences.
For example, collective activities engage the entire audience group together, and one-to-one interactions allow direct social play on the individual level---which can be fostered by affording a virtual embodiment or proxy for each individual. Live stream viewers also have a more nuanced social need of seeking social status among the audience~\cite{liExaminingGiftingBehavior2021,liSystematicReviewLiterature2020}, which we categorized as one-to-many, self-expressing interactions.
For the sake of presentation clarity, the initial design is not reported dedicatedly, but in combination with the iterations mentioned in the following subsections.

\subsubsection{Iterative Design Studies}\label{sec:iter-procedure}

After we settled the features and developed the initial MRSLS prototype, we iterated the prototype with two design studies (\cref{fig:design-process} \ding{194}).
We recruited two groups of 15 and 14 Chinese participants, respectively, aged 18--27, through social media advertisements for two rounds of qualitative studies.
The first-round participants (A1--A15) consisted of nine females and six males,
six of whom had lived in Hangzhou for more than half a year.
The second-round participants (B1--B14) included six female and eight male, and two had lived in Hangzhou for more than half a year.
\edit{Their demographics are listed in \cref{tab:iterative-demographics}.}

\begin{table}[ht]
\caption{The demographics of the iterative study participants. In Mann-Whitney U tests, the two groups show no significant differences in their familiarity
with the West Lake (\(p=0.153\)) or local culture (\(p=0.064\)). (*: 1=lowest, 7=highest)}\label{tab:iterative-demographics}
\small
\begin{tabular}[t]{p{1cm}p{1cm}p{1.5cm}ll}
\toprule
\multirow{2}{*}{ID} & \multirow{2}{*}{Age} & \multirow{2}{*}{Gender} & \multicolumn{2}{c}{Familiarity with*} \\
\cmidrule{4-5}
& & & West Lake & Local Culture \\
\midrule
A1 & 27 & Female & 5 & 5 \\
A2 & 21 & Male & 7 & 6 \\
A3 & 24 & Male & 6 & 5 \\
A4 & 23 & Female & 5 & 5 \\
A5 & 22 & Female & 5 & 5 \\
A6 & 21 & Male & 4 & 6 \\
A7 & 20 & Female & 3 & 2 \\
A8 & 22 & Female & 2 & 1 \\
A9 & 20 & Female & 3 & 3 \\
A10 & 18 & Female & 3 & 4 \\
A11 & 21 & Male & 3 & 5 \\
A12 & 20 & Male & 2 & 3 \\
A13 & 21 & Male & 5 & 5 \\
A14 & 21 & Female & 3 & 5 \\
A15 & 20 & Female & 3 & 5 \\
\midrule
B1 & 20 & Female & 3 & 3 \\
B2 & 22 & Male & 4 & 4 \\
B3 & 20 & Male & 5 & 5 \\
B4 & 26 & Female & 2 & 2 \\
B5 & 21 & Male & 2 & 3 \\
B6 & 19 & Male & 4 & 4 \\
B7 & 22 & Female & 2 & 3 \\
B8 & 19 & Female & 3 & 3 \\
B9 & 22 & Male & 1 & 3 \\
B10 & 25 & Female & 5 & 4 \\
B11 & 22 & Female & 4 & 6 \\
B12 & 21 & Male & 5 & 4 \\
B13 & 21 & Male & 2 & 3 \\
B14 & 21 & Male & 1 & 3 \\
\bottomrule
\end{tabular}
\end{table}

In both rounds, participants joined the live stream simultaneously on a designated date and time. They freely experienced the MRSLS together for 20 minutes.
To best simulate real-world live-streaming scenarios where viewers join and leave freely, participants were not provided prior feature instructions. All the instructions were displayed on the top-right corner as part of the MRSLS content.
They were neither informed that their gifts would be reimbursed later, but only told to send gifts voluntarily and similarly to their regular habits.

After the experience session, participants were invited to separate semi-structured interviews. Each received a stipend of 80 CNY plus the amount of money they spent on gifts.
The interview was mainly concerning the clarity and richness of the local culture, the visual harmony between the MR content and the scene, as well as the overall interactional user experience.
We conducted thematic analyses to extract data from the interview transcripts.

\edit{As an overview of the iterations}, the first-round participants highlighted the need for a stronger sense of immersion in the heritage site, more stylized virtual embodiment, and more salient social cues.
With feature adjustments accordingly, we confirmed that no major issues persisted in the second round, and finalized the feature design at this round.
\edit{Apart from the features, we also refined the visual appeal of the MR elements (such as the umbrella, the fish, and the lotus) according to participants' comments. To uphold interactive affordance and users' action-awareness, we did not strive to achieve an exceptionally high-fidelity and realistic style that completely blends into the scene. Nevertheless, we confirmed to have achieved satisfactory aesthetics and visual harmony  based on feedback from second-round participants.}

The following subsection details all features of our MRSLS prototype, each with the initial design, relevant findings generated from the iteration studies, and how it was iterated.

\subsection{MRSLS Prototype}\label{sec:features}

\begin{figure*}[ht]
  \centering
  \begin{subfigure}[B]{0.6\textwidth}
    \centering
    \includegraphics[width=\textwidth]{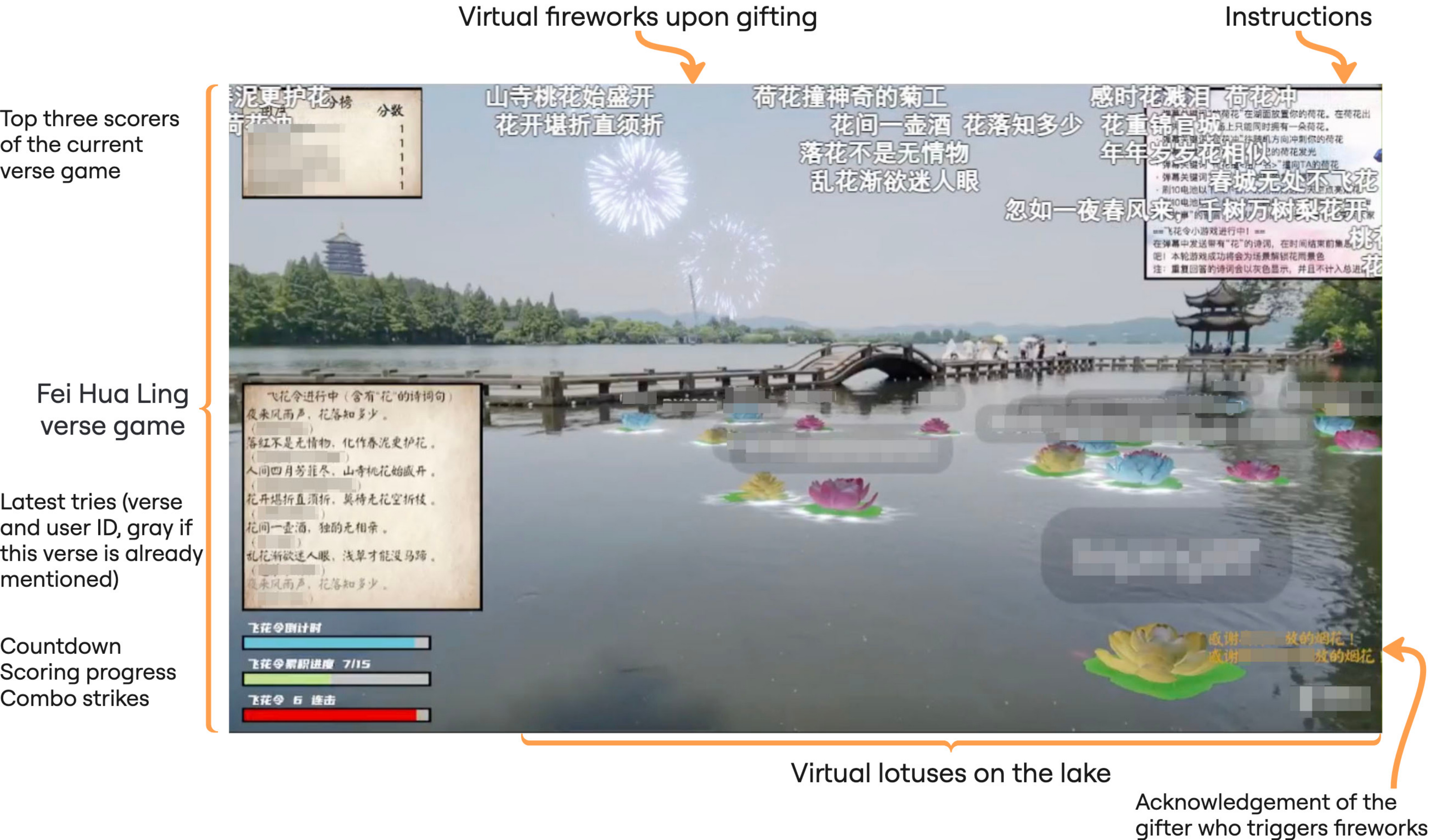}
    \Description{A moment of the West Lake scene. In addition, there are some virtual fireworks exploding in the sky. There are some virtual lotuses floating on the water with ripples and name tags. Some game UI is overlaid on the left. The instructions are overlaid on the top right.}
    \caption{A screenshot in the middle of the session. The plain scenery of regular scenic live streaming is enhanced by virtual and real-time interactive elements.}%
    \label{fig:design:features}
  \end{subfigure}\hfill
  \begin{subfigure}[B]{0.35\textwidth}
    \centering
    \includegraphics[width=0.7\textwidth]{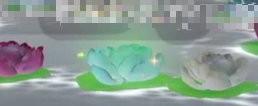}
    \Description{Three lotuses in red, blue, and white, respectively, are floating on the lake. The blue one is shining.}
    \vspace{0.2cm}\par
    \includegraphics[width=0.7\textwidth]{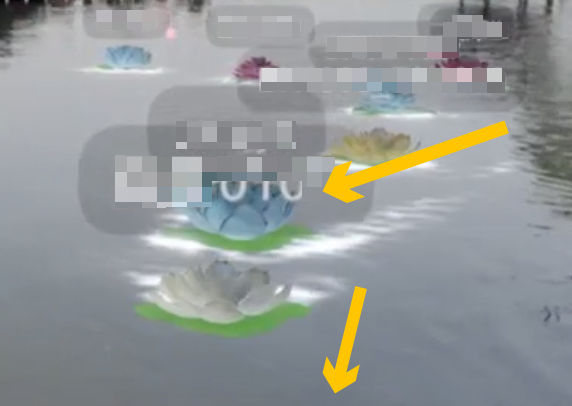}
    \caption{The lotuses on the lake, each controlled by one viewer. The lotus above is shining. The blue lotus below is dashing and colliding with the white one.}\label{fig:design:lotus}
  \end{subfigure}
  \begin{subfigure}[T]{0.25\textwidth}
    \centering
    \includegraphics[width=0.9\textwidth]{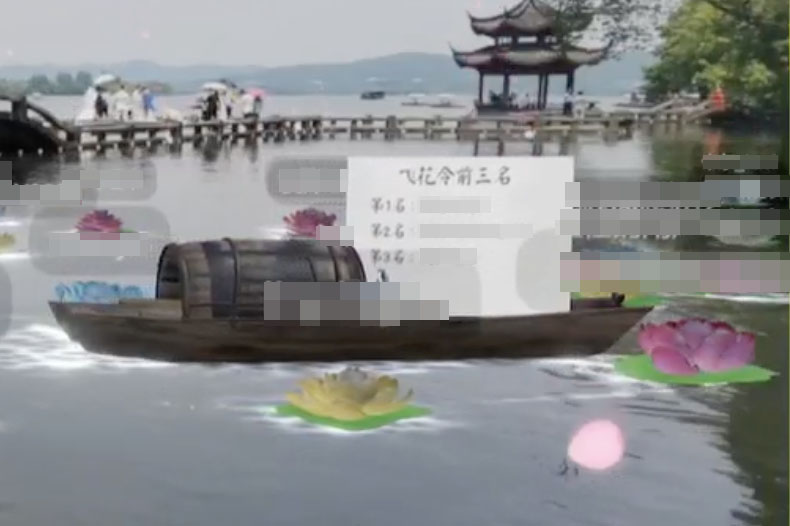}
    \Description{A boat carrying a board with the top three scorers' names is floating on the lake.}
    \caption{The boat carrying top scorers' names after the verse game.}\label{fig:design:fhl}
  \end{subfigure}\hfill
  \begin{subfigure}[T]{0.2\textwidth}
    \centering
    \includegraphics[width=\textwidth]{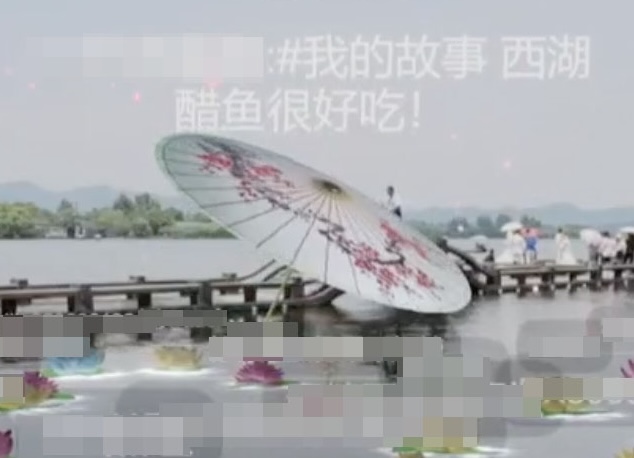}
    \Description{An oil-paper umbrella is floating in the sky. A short piece of text is above the umbrella.}
    \caption{An oil-paper umbrella with hashtagged comment.}%
    \label{fig:design:umbrella}
  \end{subfigure}\hfill
  \begin{subfigure}[T]{0.22\textwidth}
    \centering
    \includegraphics[width=0.8\textwidth]{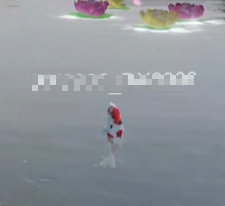}
    \Description{A fish with a name tag jumping out of water.}
    \caption{A fish jumping out of water.}
    \label{fig:design:fish}
  \end{subfigure}\hfill
  \begin{subfigure}[T]{0.2\textwidth}
    \centering
    \includegraphics[width=0.6\textwidth]{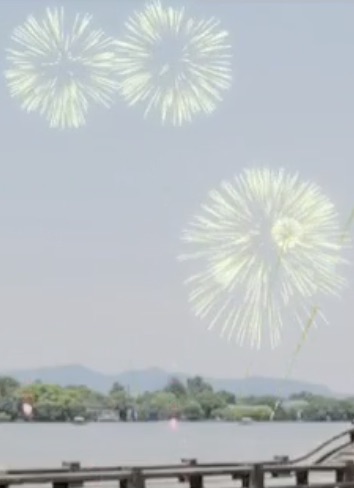}
    \Description{Some yellow fireworks exploding in the sky.}
    \caption{Fireworks on the sky.}\label{fig:design:firework}
  \end{subfigure}
  \caption{The MRSLS design. Viewers can actively engage in interactions with the scene. The features aim to recreate in-person collective activities related to local cultures.}\label{fig:iteration}
\end{figure*}

The screenshots of our MRSLS prototype are shown in \cref{fig:iteration}.
And the features are detailed in the following subsections.

\subsubsection{Camera View}

\begin{figure}[ht]
  \centering
  \includegraphics[width=\linewidth]{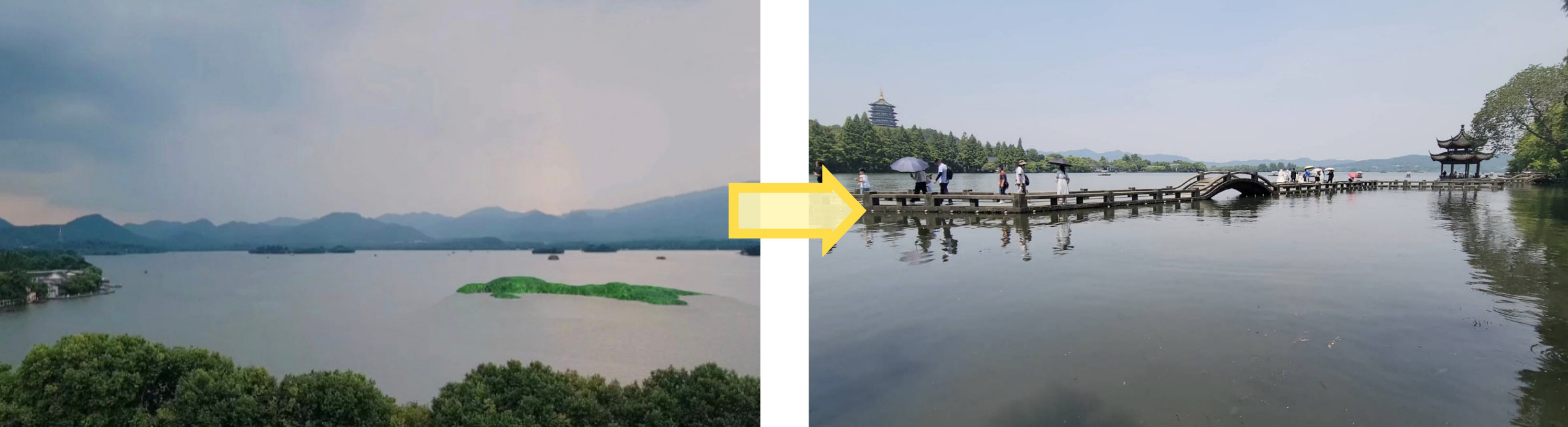}
  \Description{An arrow connects two pictures. On the left is an overlooking view from one side of the West Lake. The building and trees are far away. On the right is a first-person view of one side of West Lake. The buildings and trees are larger and clearer. The view also captures a bridge and some pedestrians.}
  \caption{The initial overlooking view (left) is preferred by plain SLS like \href{http://skylinewebcams.com}{SkylineWebcams} and \href{https://livechina.cctv.com}{Live China}. The new first-person view (right) better fits interactive MRSLS.}\label{fig:view}
\end{figure}

We initially adopted a top-down camera view that overlooks the target landscape.
It captures West Lake and the distant hills from a lookout point (\cref{fig:view}).
This choice aligned with the common preference of existing SLS platforms, such as \href{http://skylinewebcams.com}{SkylineWebcams} and \href{https://livechina.cctv.com}{Live China}, and it echoed the formative study, where some participants expressed a desire for new perspectives.

However, most participants in the first round did not find the overlooking view appealing.
Despite the new perspective, participants could hardly build connections with it or develop cultural conceptualization of the spot (A1--4, A12--14).
\textbf{The overlooking perspective may have situated the audience as spectators from a distance, instead of being contextualized at the remote site}.
In contrast, traditional SLS can effectively adopt such views because viewers typically adopt a passive spectating role.
Practically speaking, an overlooking view could neither clearly capture representative landmarks and real-world objects (A1, A3, A4, A10), leading to inconsistent sizes of real and virtual objects (A3, A11, A15).
Therefore, in the revised version, we adopted a first-person angle
at ``Orioles Singing in the Willows'' --- one of the top 10 scenes of West Lake~\cite{geRelationshipSoundscapeChinese2012}.
The view clearly included a pavilion, the Leifeng Pagoda\footnote{\url{https://en.wikipedia.org/wiki/Leifeng_Pagoda}},
and a causeway for pedestrians (\cref{fig:view}).
Participants in the second round generally acknowledged that the new camera view was realistic (B1--5, B8--10, B13--14) and immersive (B2, B4, B10, B13--B13).
The inclusion of local landmarks and architectural styles made the West Lake more recognizable (B3, B5, B7, B9).
The pedestrians also made the scene more lively and dynamic (B2--B2, B5, B9).

\subsubsection{Lotus Spreading}\label{feat:lotus-spreading}

To support direct individual (one-to-one) interactions, this feature allows every MRSLS viewer to release a lotus on the water surface with the viewer's name tagged to it, using the command ``\texttt{release lotus}''.
Subsequently, viewers can initiate amusing and lightweight mutual interactions with the command ``\texttt{dash my lotus}'', which makes the viewer's own lotus to ``dash'' in a random direction and presumably collide with others' lotuses.
To prevent virtual lotuses from occluding the scene, they slowly drift away to the right when idle and are destroyed when outside the viewport.

The purpose of this feature is threefold. First, the lotus is a common scenery object on the West Lake and is commonly appreciated in classical Chinese poetry about West Lake. It is also a cultural tradition to release lotus-shaped lanterns collectively on lakes during festivals.
Hence, virtual lotuses on the lake imbue the live stream with extra ornamental and cultural values.
Second, the one-to-one association between the viewer and the lotus, together with the autonomous mobility control, affords virtual embodiment on the remote site~\cite{genayBeingAvatarReal2022,kilteniSenseEmbodimentVirtual2012}.
The stronger sense of ``being there'' shall reinforce the connection between the SLS viewers and the streamed heritage site~\cite{gonzalezKeepingStrongConnections2007, dengBlendedTourismExperiencescape2019}.
Third, the dashing interaction enables entertaining and teasing social play between viewers~\cite{marquezseguraBodyGamesDesigning2013}, which can strengthen the social and participatory attributes of the live stream.

During the iteration studies, participants expressed their desire for \textbf{more direct cues of interpersonal interactions} and \textbf{clearer distinction and uniqueness among viewers' visual representations} --- possibly due to the strong sense of embodiment of the lotus and the social need in live streams~\cite{sjoblomWhyPeopleWatch2017,webbDistributedLivenessUnderstanding2016}.
Hence, we first added the command ``\texttt{hit <ID> with my lotus}'' to make one's lotus dash into another user's lotus like bumper cars (A9, A15).
We also increased the dash speed and bounciness of all lotuses,
enabling them to sprint toward other lotuses faster and collide more dramatically.
Then, we assigned a random color to each lotus upon creation,
including pink, white, yellow, and blue --- all of which are available in the real world (\cref{fig:design:lotus}).
Second, we added the command ``\texttt{shine my lotus}'' to trigger a shining effect on the user's lotus for two seconds (also suggested by A9--A10) (\cref{fig:design:lotus}).
The shining effect could assist users in locating their lotuses with a subtle and harmonious visual effect.
The second-round participants generally accepted the new lotus design,
with B7 and B11 particularly noting that the new lotuses are beautiful and match the environment well.

\subsubsection{Chinese Verse Game ``Fei Hua Ling''}\label{sec:features:feihua}

To deeply connect viewers with the site's rich intangible heritage---and also to foreground communal engagement, we designed a feature that transforms passive scenery appreciation into a collective act of cultural recall.
We adapted the traditional Chinese literary game ``Fei Hua Ling'' (
  \begin{CJK*}{UTF8}{gbsn}飞花令
\end{CJK*}, lit.\@ ``flying flower command''), which dates back to the Tang Dynasty (618--907).
Beyond mere entertainment, it is also a common collective ritual tradition during gathering and scenery appreciation.
In our collaborative version, the audience works together to recite classical Chinese poems related to a specific topic within a 5-minute limit, winning if the total number of unique verses exceeds a certain threshold.

The interaction was purpose-built to bridge the tangible scenery of West Lake with the vast reservoir of poetry it has inspired.
During our user studies, we launched two rounds of the game in each session, and the chosen two topics directly reinforced this link.
The first topic was ``flower'', which served as a callback to the virtual lotuses on the lake and the default topic at the game's origin.
The second topic is ``Hangzhou'' (the city which West Lake is at) or Jiangnan (
  \begin{CJK*}{UTF8}{gbsn}江南
\end{CJK*}, lit. `South of the Yangtze River', a geographic area where Hangzhou resides)'', which prompted remembrance of the site itself.
By simultaneously viewing the landscape and reciting the poetry associated with it, participants could engage in a culturally immersive and imaginative experience, layering the live stream with historical and literary depth.
This collective effort also served to enhance the sense of active, macro-level participation in the stream.

As participants indicated the need for stronger cues of interpersonal interactions, A8 and A9 suggested increasing the competitiveness of the game.
Therefore, we added a scoreboard that displays the current top three scorers in real time (\cref{fig:design:features}),
and a huge virtual boat crossing the lake at the end of the game with the top three IDs on a flag and absurdly breaking away all the lotuses on its course (\cref{fig:design:fhl}).

\subsubsection{Fish Feeding}\label{sec:features:fish}

As a simple and introductory feature for the entire MRSLS prototype and concept,
viewers can send the command ``\texttt{feed fish}'' to trigger a one-time animation of throwing fish feed at a random location on the lake, followed by a local ornamental fish jumping out of water and a water splashing sound.
Although the feature is straightforward, it gives viewers the ability to add to the scene's ornamental values and dynamism.

But during the iteration study, the inability to identify which participant triggered which fish-feeding animation discouraged them from using this feature again (A3--A4, A9, A11).
The innate delay of live stream footage also amplified this issue.
This reveals that \textbf{live stream viewers often expect immediate confirmations of their actions and links between them and the live stream content}, even if the actions bear no embodiment or profound reciprocal meaning. Therefore, we attached a name tag to each fish that follows its movement (\cref{fig:design:fish}), similar to that of a lotus. With the name tags, the issue was resolved in the second round, and B9 explicitly mentioned that the name tags were useful in confirming their interactions with the system.

\subsubsection{Fireworks and Oil-Paper Umbrella with Stories}

Another common expectation in live streams is to seek self-expression and involvement in the streamed visual content through gifting and the streamer's subsequent acknowledgement~\cite{luYouWatchYou2018, kimWhoWillSubscribe2019,zhouMagicDanmakuSocial2019,leeUnderstandingHowDigital2019}.
This nuanced social factor transforms gifting into what we categorized as ``one-to-many'' interactions in \cref{sec:initial-design}.
To compensate for the lack of the possibility in SLS, we introduced two gift-exclusive interactions with different price thresholds.
A gift below 10 CNY will trigger a virtual fireworks animation in the sky with the gifter's name displayed (\cref{fig:design:firework}).
A gift above 10 CNY will allow the viewer to share a personal story in the live chat with ``\texttt{\#MyStory}'' hashtag, which will be highlighted and displayed with an oil-paper umbrella flying across the screen (\cref{fig:design:umbrella}).

Both features also take into account visual harmony and cultural embedding. Oil-paper umbrellas are iconic, national heritage-level local crafts~\cite{OilPaperUmbrellaMaking} with the symbolism of gathering and reunion~\cite{renRedesignTraditionalChinese2013}. Fireworks are also an ancient Chinese invention, and setting off fireworks is believed to promise luck~\cite{timothys.y.lammuseumofanthropologyChineseNewYear2021}.
Participants have not raised particular concerns about these two features, hence they are kept intact during the iteration.

\subsection{Technical Implementation}

\begin{figure}[ht]
  \centering
  \includegraphics[width=0.8\linewidth]{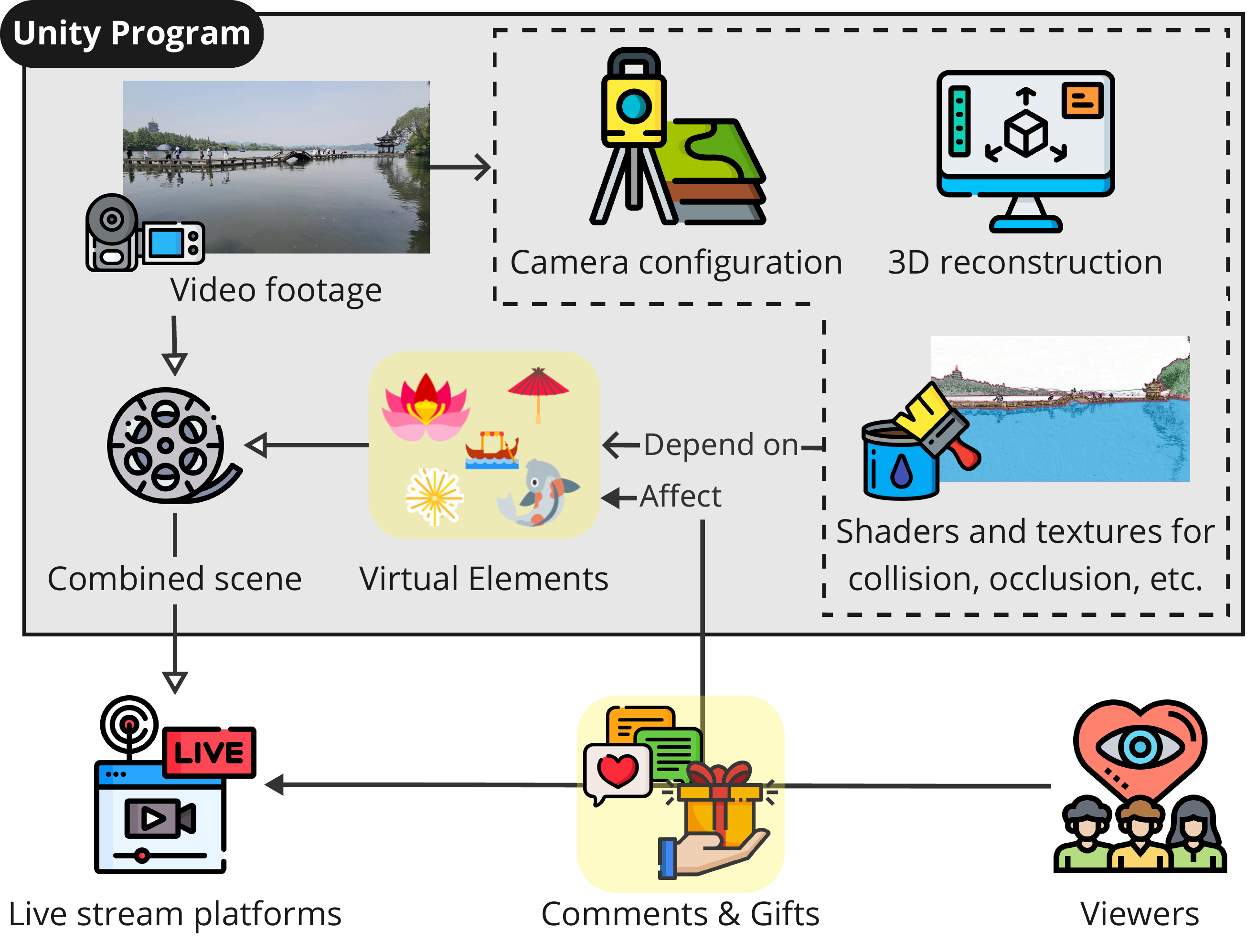}
  \Description[A diagram of the MRSLS system implementation.]{A video stream from the camera is fed to a Unity program, which contains a “real-time” part and a “one-time” part. The one-time part uses the footage to calculate camera configurations and 3D reconstruction. The real-time part renders (1) the footage with proper camera configurations, (2) materials, shaders, and textures for collision and occlusion thanks to the 3D reconstruction, and (3) the MR elements that depend on the materials, shaders, and textures. The combined scene of the entire real-time part is sent to live-streaming platforms. Viewers' activities (comments and gifts) are sent to both the live-streaming platform and the Unity program to control the MR elements.}
  \caption{The software architecture of our MRSLS prototype. Icons in this figure are designed by Freepik, Smashicons, Icons8, and Eucalyp.}\label{fig:architecture}
\end{figure}

As shown in \cref{fig:architecture}, the live stream view is the real-time rendering output of a Unity program. The Unity scene's camera is configured according to the physical camera that captures the scenic view. Then, proper 3D reconstructions are made to describe the geographical structure of the view with transparent collision boundaries, and virtual elements are added according to them. The Unity program sets up real-time network communication with the live streaming platform for viewers' comments and gifts, which further control the virtual elements.

The implementation was in collaboration with a local company, \company, who developed the network infrastructure and the 3D assets such as the shader and texture for selective occlusion, and the lake boundary with ripple hitbox effects.

\section{Evaluation}

In a between-subjects, mixed-methods user study, we evaluated the West Lake MRSLS prototype in parallel with a plain SLS of the same content.
In line with the main focuses of our design process, we primarily aim to answer the following research questions (RQs):
\begin{enumerate}[label=\textbf{RQ\arabic*}]
  \item Can properly designed MRSLS strengthen the connection between the audience and the live-streamed heritage site?\label{RQ:scene}
  \item Can properly designed MRSLS reinforce a participatory and social experience that further engages the audience with the heritage site?\label{RQ:engage}
  \item How does the audience perceive and position MR interactivity in SLS\@ in general?\label{RQ:interact}
\end{enumerate}

\subsection{Participants and Procedures}\label{sec:protocol}

Participants were recruited through social media and word-of-mouth.
Each participant was assigned a condition (MRSLS or SLS) upon signing up. The plain SLS condition removed all the interactive features and the virtual objects, leaving only the camera footage.
For either condition's participant group, they were instructed to enter the specific live stream room on a specific date and time together and freely use it for 20 minutes.
We recruited equal numbers of participants for the two conditions; but after excluding the no-shows, there were 43 participants (namely, P1--P43) for MRSLS and 36 participants (namely, Q1--Q36) for SLS.
Participants' demographic information is shown in \cref{tab:demographics}.

\begingroup
\setlength{\tabcolsep}{3.5pt} 
\begin{table*}[t]
\caption{The demographics of the main user study. In Mann-Whitney U tests, the two groups show no significant differences in their familiarity with the West Lake (\(p=0.156\)) or the local culture (\(p=0.378\)). (*: 1=lowest, 7=highest)}\label{tab:demographics}
\small
\begin{tabular}[t]{lllll}
\toprule
\multirow{2}{*}{ID} & \multirow{2}{*}{Age} & \multirow{2}{*}{Gender} & \multicolumn{2}{c}{Familiarity with*} \\
\cmidrule{4-5}
& & & {\footnotesize West Lake} & {\footnotesize Local Culture} \\ 
\midrule
P1  & 27 & Male & 2 & 1\\ 
P2  & 21 & Female & 4 & 4\\ 
P3  & 21 & Male & 3 & 3\\ 
P4  & 26 & Female & 4 & 3\\ 
P5  & 23 & Male & 6 & 6\\ 
P6  & 21 & Male & 1 & 1\\ 
P7  & 22 & Female & 1 & 2\\ 
P8  & 21 & Male & 3 & 3\\ 
P9  & 24 & Female & 5 & 3\\ 
P10 & 19 & Male & 6 & 3\\ 
P11 & 25 & Female & 2 & 2\\ 
P12 & 27 & Male & 3 & 2\\ 
P13 & 24 & Male & 4 & 5\\ 
P14 & 23 & Female & 3 & 2\\ 
P15 & 20 & Male & 5 & 3\\ 
P16 & 25 & Male & 2 & 2\\ 
P17 & 20 & Undisclosed & 1 & 1\\ 
P18 & 24 & Female & 5 & 5\\ 
P19 & 26 & Male & 5 & 4\\ 
P20 & 21 & Male & 3 & 3\\ 
P21 & 24 & Male & 2 & 1\\ 
P22 & 21 & Female & 1 & 1\\ 
P23 & 25 & Female & 3 & 4\\ 
P24 & 23 & Female & 3 & 2\\ 
P25 & 29 & Male & 6 & 4\\ 
P26 & 21 & Female & 2 & 2\\ 
P27 & 21 & Female & 3 & 3\\ 
\bottomrule
\end{tabular}
\hfill
\begin{tabular}[t]{lllll}
\toprule
\multirow{2}{*}{ID} & \multirow{2}{*}{Age} & \multirow{2}{*}{Gender} & \multicolumn{2}{c}{Familiarity with*} \\
\cmidrule{4-5}
& & & {\footnotesize West Lake} & {\footnotesize Local Culture} \\ 
\midrule
P28 & 24 & Male & 3 & 2\\ 
P29 & 20 & Male & 3 & 2\\ 
P30 & 21 & Female & 4 & 4\\ 
P31 & 26 & Male & 7 & 6\\ 
P32 & 19 & Female & 3 & 4\\ 
P33 & 22 & Male & 6 & 5\\ 
P34 & 24 & Non-binary & 7 & 4\\ 
P35 & 24 & Male & 3 & 3\\ 
P36 & 23 & Female & 5 & 6\\ 
P37 & 23 & Male & 5 & 3\\ 
P38 & 22 & Female & 4 & 4\\ 
P39 & 27 & Female & 4 & 3\\ 
P40 & 22 & Male & 1 & 1\\ 
P41 & 21 & Male & 1 & 1\\ 
P42 & 24 & Male & 6 & 5\\ 
P43 & 20 & Male & 5 & 2\\ 
\midrule
Q1  & 20 & Male & 1 & 1\\ 
Q2  & 27 & Male & 4 & 2\\ 
Q3  & 24 & Female & 3 & 3\\ 
Q4  & 24 & Female & 3 & 2\\ 
Q5  & 25 & Female & 4 & 3\\ 
Q6  & 25 & Male & 3 & 3\\ 
Q7  & 24 & Female & 1 & 1\\ 
Q8  & 23 & Female & 1 & 1\\ 
Q9  & 24 & Female & 2 & 2\\ 
Q10 & 23 & Male & 3 & 3\\ 
\bottomrule
\end{tabular}
\hfill
\begin{tabular}[t]{lllll}
\toprule
\multirow{2}{*}{ID} & \multirow{2}{*}{Age} & \multirow{2}{*}{Gender} & \multicolumn{2}{c}{Familiarity with*} \\
\cmidrule{4-5}
& & & {\footnotesize West Lake} & {\footnotesize Local Culture} \\ 
\midrule
Q11 & 21 & Male & 1 & 1\\ 
Q12 & 25 & Male & 4 & 4\\ 
Q13 & 21 & Female & 4 & 1\\ 
Q14 & 25 & Male & 1 & 1\\ 
Q15 & 21 & Male & 1 & 3\\ 
Q16 & 30 & Male & 2 & 1\\ 
Q17 & 25 & Female & 4 & 4\\ 
Q18 & 25 & Female & 3 & 3\\ 
Q19 & 23 & Male & 4 & 3\\ 
Q20 & 30 & Female & 6 & 5\\ 
Q21 & 25 & Female & 2 & 1\\ 
Q22 & 30 & Male & 3 & 2\\ 
Q23 & 23 & Female & 3 & 2\\ 
Q24 & 18 & Male & 4 & 4\\ 
Q25 & 26 & Male & 5 & 5\\ 
Q26 & 30 & Female & 2 & 4\\ 
Q27 & 22 & Male & 1 & 1\\ 
Q28 & 25 & Female & 5 & 3\\ 
Q29 & 20 & Female & 6 & 6\\ 
Q30 & 20 & Female & 3 & 3\\ 
Q31 & 26 & Male & 2 & 2\\ 
Q32 & 25 & Male & 3 & 3\\ 
Q33 & 19 & Female & 2 & 2\\ 
Q34 & 25 & Male & 4 & 4\\ 
Q35 & 23 & Male & 5 & 5\\ 
Q36 & 27 & Male & 1 & 1\\ 
\bottomrule
\end{tabular}
\end{table*}
\endgroup

When both groups were invited to their designated live streams, they were told to freely experience together with other audience members for 20 minutes, with the same protocol as the previous iteration studies (\cref{sec:iter-procedure}).
After experiencing the live stream, participants were invited to complete a questionnaire.
Same as the previous studies, each participant received a stipend of 80 CNY, together with the amount of money they spent on gifts.

Considering MRSLS's unique fusion of site appreciation and live-streaming entertainment, the questionnaire was constructed based on several existing measurements, including ``place attachment''~\cite{raymondMeasurementPlaceAttachment2010}, affective and cognitive ``immersion'' in entertaining systems~\cite{jennettMeasuringDefiningExperience2008}, gamified information retrieval about travel destinations~\cite{sigalaApplicationImpactGamification2015}, synchronous colocated XR games~\cite{bhattacharyyaBrickModelDesigning2019}.
We cherry-picked relevant items from their metrics, and modified them to fit our context of ``online, real-time, and collective experience of a cultural heritage site''.
In addition, we also added some items based on the comments in previous iterative tests.
In the questionnaire, we included not only quantitative ratings in a 7-point Likert scale (0: strongly disagree; 6: strongly agree), but also qualitative data from open-ended questions about participants' reasoning behind their ratings.
In addition, the MRSLS group received extra questions about the design of the MR elements.
The questions shared by both groups were compared using Mann-Whitney U tests.

\section{Findings}\label{sec:eval}

The general statistical comparison results are listed in \cref{tab:mwu}, and the following subsections detail the findings.
\Cref{sec:eval:place} looks into viewers' connection with the live-streamed place through the MRSLS experience (\ref{RQ:scene}).
\Cref{sec:eval:engage} then elaborates the increment in the sense of participation and engagement (\ref{RQ:engage}).
Finally, \cref{sec:eval:ar} examines the general appropriateness of the MR design, which serves as a lens to participants' mental positioning of MRSLS experience at heritage sites (\ref{RQ:interact}).

\subsection{Place Connection}\label{sec:eval:place}

When designing MRSLS as a heritage site interaction, we are most concerned with viewers' connections with the heritage site.
It is similar to how some migrants watch plain SLS of their hometowns to regain attachments with them~\cite{gonzalezKeepingStrongConnections2007,jarrattExplorationWebcamtravelConnecting2021},
but goes beyond the common notion of ``place attachments'', which emphasizes the identity, dependence, and bonding with the place as a resident or former resident~\cite{raymondMeasurementPlaceAttachment2010}.
To examine similar effects on the general audience, we first decomposed ``place'' as a concept that interweaves the physical space and people's interpretations and perceived meanings of it~\cite{aucoinAnthropologicalUnderstandingSpace2017}.
Accordingly, we distinguished two major foci on participants' MRSLS experiences: the sense of situatedness (``being there''), and the recognition of the place's values.

The questionnaire shows that MRSLS participants have gained a stronger sense of ``being there'', and a stronger interest in the place (\cref{tab:mwu}).
This validates our intentions of affording virtual embodiments that situate themselves at the site and roam around using the lotus feature.
As participants further elaborated, the plain SLS sometimes created a distant feeling (Q12, Q17), and viewers might not always sense connections with the place (Q22). Q17 articulated a sense of \emph{``peeking in, without close and real-time engagement with what's happening there''}.
Q12 added, \emph{``the happiness belongs to them (the pedestrians at the site). I'm just watching from a distance''}.

Apart from the sense of ``being there'', the MR features also raised the values of the live-streamed place in various aspects. The objects themselves added extra ornamental values to the view (P6, P39); they also made the scene more lively and dynamic (P34), as fixed-camera scenery view is sometimes too still and boring (11 out of 37).
Besides, the cultural elements communicated through the representative MR objects and the poetical game deepened the viewer experience as well (P20, P25).

\begin{table*}[t]
  \caption{Mann-Whitney U test results of the questionnaire feedback (excluding MRSLS-only questions). The effect sizes and powers of non-significant rows (p-value \(<0.05\)) are omitted.}
  \label{tab:mwu}
  \footnotesize\begin{tabularx}{\textwidth}{Xllllll}
  \toprule
  Question & \makecell[l]{SLS mean (std)} & \makecell[l]{MRSLS mean (std)} & U-value & p-value & Effect size & Power \\
  \midrule
  \S\;\textit{Place Connection}\\
  The view gives me a sense of “being there” & 2.80 (1.61) & 3.98 (1.30) & 445.5 & \textbf{0.0017} & 0.30 & 0.36 \\
  \addlinespace[\aboverulesep]
  This experience has recalled my previous memories with this place (if applicable) & 3.95 (1.60) & 4.13 (1.43) & 301.0 & 0.7890 & 0.48 & --- \\
  \addlinespace[\aboverulesep]
  This experience has triggered my interest in this place (if applicable) & 2.77 (1.37) & 3.58 (1.34) & 312.5 & \textbf{0.0234} & 0.33 & 0.36 \\
  \midrule
  \S\;\textit{Participatory and social experience}\\
  This experience provides a sense of community & 3.57 (1.54) & 4.35 (1.15) & 558.0 & \textbf{0.0425} & 0.37 & 0.49 \\
  \addlinespace[\aboverulesep]
  I have had direct communications/interactions with other viewers & 2.97 (2.04) & 3.42 (1.80) & 659.0 & 0.3416 & 0.44 & --- \\
  \addlinespace[\aboverulesep]
  I am motivated to type comments in this live stream & 3.66 (1.57) & 4.44 (1.37) & 519.0 & \textbf{0.0159} & 0.34 & 0.44 \\
  \addlinespace[\aboverulesep]
  I am motivated to send gifts in this live stream & 2.23 (1.65) & 3.09 (1.56) & 535.0 & \textbf{0.0267} & 0.36 & 0.46 \\
  This experience provides a comparable level of engagement with games & 1.91 (1.22) & 3.58 (1.52) & 300.5 & \textbf{0.0000} & 0.20 & 0.22 \\
  \midrule
  \S\;\textit{General UX}\\
  I am satisfied about this live steam experience & 2.71 (1.54) & 4.21 (1.28) & 344.5 & \textbf{0.0000} & 0.23 & 0.26 \\
  \addlinespace[\aboverulesep]
  I want to experience this live stream again & 2.40 (1.65) & 4.05 (1.45) & 345.5 & \textbf{0.0000} & 0.23 & 0.26 \\
  \addlinespace[\aboverulesep]
  This experience is interesting & 2.51 (1.52) & 4.56 (1.22) & 207.0 & \textbf{0.0000} & 0.14 & 0.16 \\
  \addlinespace[\aboverulesep]
  This experience is appropriate & 2.60 (1.29) & 4.28 (1.28) & 276.0 & \textbf{0.0000} & 0.18 & 0.21 \\
  \addlinespace[\aboverulesep]
  This experience is immersive & 2.63 (1.83) & 4.00 (1.27) & 430.0 & \textbf{0.0010} & 0.29 & 0.35 \\
  \addlinespace[\aboverulesep]
  The interactions in this experience is easy & 3.00 (1.85) & 4.49 (1.37) & 402.5 & \textbf{0.0004} & 0.27 & 0.32 \\
  \addlinespace[\aboverulesep]
  The view is aesthetically pleasant & 3.54 (1.74) & 4.07 (1.47) & 621.0 & 0.1728 & 0.41 & --- \\
  \addlinespace[\aboverulesep]
  The view is vivid and lively & 2.97 (1.58) & 4.07 (1.52) & 460.0 & \textbf{0.0028} & 0.31 & 0.38 \\
  \addlinespace[\aboverulesep]
  The view reflects elements unique to this place & 3.17 (1.64) & 4.16 (1.27) & 500.0 & \textbf{0.0089} & 0.33 & 0.42 \\
  \addlinespace[\aboverulesep]
  Overall, I like the view of this live stream & 3.29 (1.58) & 4.26 (1.14) & 505.5 & \textbf{0.0102} & 0.34 & 0.43 \\
  \bottomrule
\end{tabularx}

\end{table*}

Lastly, we noticed that for those participants already possessing personal experiences with the West Lake, the MRSLS does not significantly trigger more memories compared to plain SLS (\cref{tab:mwu}).
Q19's comment echoed this point: \emph{``when I am nostalgic, I want to quietly appreciate my hometown's view, as if I am back home. At that moment, any interaction can be a disturbance.''}
As previous literature has already identified plain SLS's benefit of relieving nostalgia and homesickness~\cite{gonzalezKeepingStrongConnections2007,jarrattExplorationWebcamtravelConnecting2021}, it may bring only a marginal gain in reinforcing place connections through MR interactions' amplified cultural imprint.
Nevertheless, MRSLS can still benefit them with novel perspectives of a familiar place (P25, P34).
\emph{``It is a much unexpected experience to have such interactions with a place I am quite familiar with,''} said P25, a college student in Hangzhou, \emph{``and I suppose I would have felt even more strongly if I had graduated and left the city.''}
P34 also affirmed: \emph{``Although I have watched some [non-MR, non-scenic] interactive streams before, the content here is significantly different. The traditional culture is more impressive.''}

\subsection{Participatory and Social Experience}\label{sec:eval:engage}

As Hamilton et al.\@ point out, shared activities in live streams deliver ``sociability'', which stresses the sense of association through the ``sheer pleasure of being together''~\cite{hamiltonStreamingTwitchFostering2014}.
According to \cref{tab:mwu}, MRSLS provides stronger motivations for viewers to initiate interactions and a stronger sense of community.
In particular, participants validated the proxied social play through metaphorical ``nudging'' and ``tickling'' of the lotus feature (P6, P8, P30, P35--36, P41, P43), as well as how the verse game offers collective engagement and topics of discussions (P9, P28, P35).
P9 said, \emph{``I am impressed by Fei Hua Ling. It is fun to play, and people have been actively participating in the game. They discussed a lot as well.''}
P28 added, \emph{``The interactions are quite interesting, and they do bring me closer to other viewers.''}
Meanwhile, 27\% plain SLS participants have explicitly complained about the lack of engaging interpersonal interactions.
Q21 commented, \emph{``The audience is watching the live stream like mere spectators. Thus, this live stream has the same effect as an online video, losing the necessity of the particular form of live streaming.''}

Despite the stronger sense of community mentioned above, MRSLS viewers tend to maintain playful and implicit interactions instead of profound communications.
They did not report significantly more direct communications with others (\cref{tab:mwu}).
We also observe limited serious use of the umbrella storytelling feature, as participants have mostly exploited this feature as fancier live comments (P2, P36).

Another interesting phenomenon is that, though acknowledging the sense of community in MRSLS, some introverted participants who do not like live streams also disliked interpersonal MR features (P18, P24, P31, P38).
As an example, P18 said, \emph{``The worst experience was when my lotus was pushed away\dots{} The users were too random, so I don't want to interact with them. I prefer just playing games myself.''}
From a very special perspectives, we highlight that such feedback also verifies the improved level of participation and social association, making it on par with other live stream genres.
Nonetheless, as live streams in general are not a multimedia experience suitable for everyone, MRSLS is neither intended to completely replace plain SLS. Rather, we intend to provide an alternative heritage site interactive experience with better alignment with other live stream experiences and richer collective cultural activities.

\subsection{The MR Design: Visual and Functional}\label{sec:eval:ar}

\begin{figure*}[ht]
  \centering
  \includegraphics[width=0.8\textwidth]{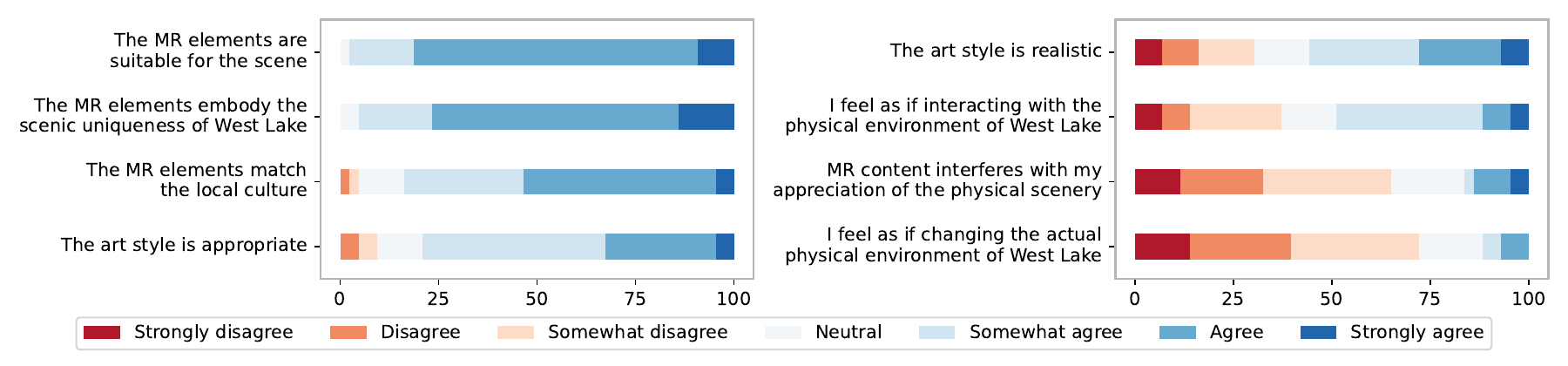}
  \caption{Questionnaire responses on MR design}
  \Description{A Likert scale plot of user feedback on MR design. Almost all participants agree that “The MRSLS content design is appropriate”, “The MR elements are suitable for the scene”, “The MR elements embody the scenic uniqueness of West Lake”, “The MR elements match the local culture”, and “The art style is appropriate”. Slightly more than half of the participants think that “The art style is realistic” and “I feel as if interacting with the physical environment of West Lake”. Almost all participants DISAGREE that “I feel as if I changed the physical environment of West Lake”.}
  \label{fig:eval:design}
\end{figure*}

\Cref{tab:mwu} shows that the overall user experience of the MRSLS was significantly better than the plain SLS, with the most notable differences in higher satisfaction (\(p<.0001\)), interest (\(p<.0001\)), and desire to use the system again (\(p<.0001\)).
In terms of the visual perception of the view, the MR content wa considered to have improved vibrancy and uniqueness, while retaining a similar level of aesthetic values.
\Cref{fig:eval:design} furthermore presents responses to the questions specific about MRSLS, showing the appropriateness of the design, its harmony with the scene, and distinguishability from the real world.

Meanwhile, participants held distinctive opinions on whether the current MR design was consistent or complementary with the real world.
Regarding the visual style, around half of MRSLS participants thought it was \emph{realistic}-looking.
For example, P39 remarked, \emph{``The occasional (virtual) fish jumping out of water and their splashing sounds have added aesthetic values to the scene.''}
On the contrary, the opposing voices thought it was cartoonish, which was not necessarily a downside: \emph{``I especially laughed at the fish jumping out of the water; they looked absurd,''} said P34.

Such distinctions, together with different visions of future MRSLS, allow us to use our MRSLS prototype as a probe that discover participants' different positions on its use purposes and style focuses.
Some participants underlined the use case of \textbf{scenic appreciation} as a more vivid SLS.
In this case, most participants preferred an \textbf{authentic style}, calling for immersive (11 out of 43) and harmonious (P6, P9, P28, P35, P37) visual design and a cautious, restrained choice of interactions (P27, P30).
Some participants still preferred an \textbf{imaginative style}, which would advocate an explicit visual distinguishment from reality (P18, P24, P31) and wider interactions (P8, P13) that really fostered a vibrant atmosphere (P3, P12, P14, P31).
Some participants would rather turn to MRSLS for \textbf{entertainment}. In this case, possible interactions that follow the \textbf{authentic style} could imitate real-world on-site activities, such as taking photos of viewers' virtual check-in (P7) and reading the site's introductions (P28), or stay non-intrusive to the site, such as gathering collectibles (P37).
In contrast, those interactions in the \textbf{imaginative style} could be playful and omnipotent, such as changing the weather (P13).

\section{Discussion}\label{sec:discussion}

Given our proposal of the MRSLS medium, an exemplar MRSLS prototype, and its careful design process, we would like to reiterate that MRSLS is not intended to replace plain SLS.
Rather, it should be seen as an alternative experience with richer interactions and vibrancy.
In this section, we further extract insights and design implications for MRSLS, propose directions for future research, and acknowledge this work's limitations.

\subsection{Unique Affordances of MR for Scenic Live Streams}

\begin{figure}[h!]
  \centering
  \includegraphics[width=0.55\linewidth]{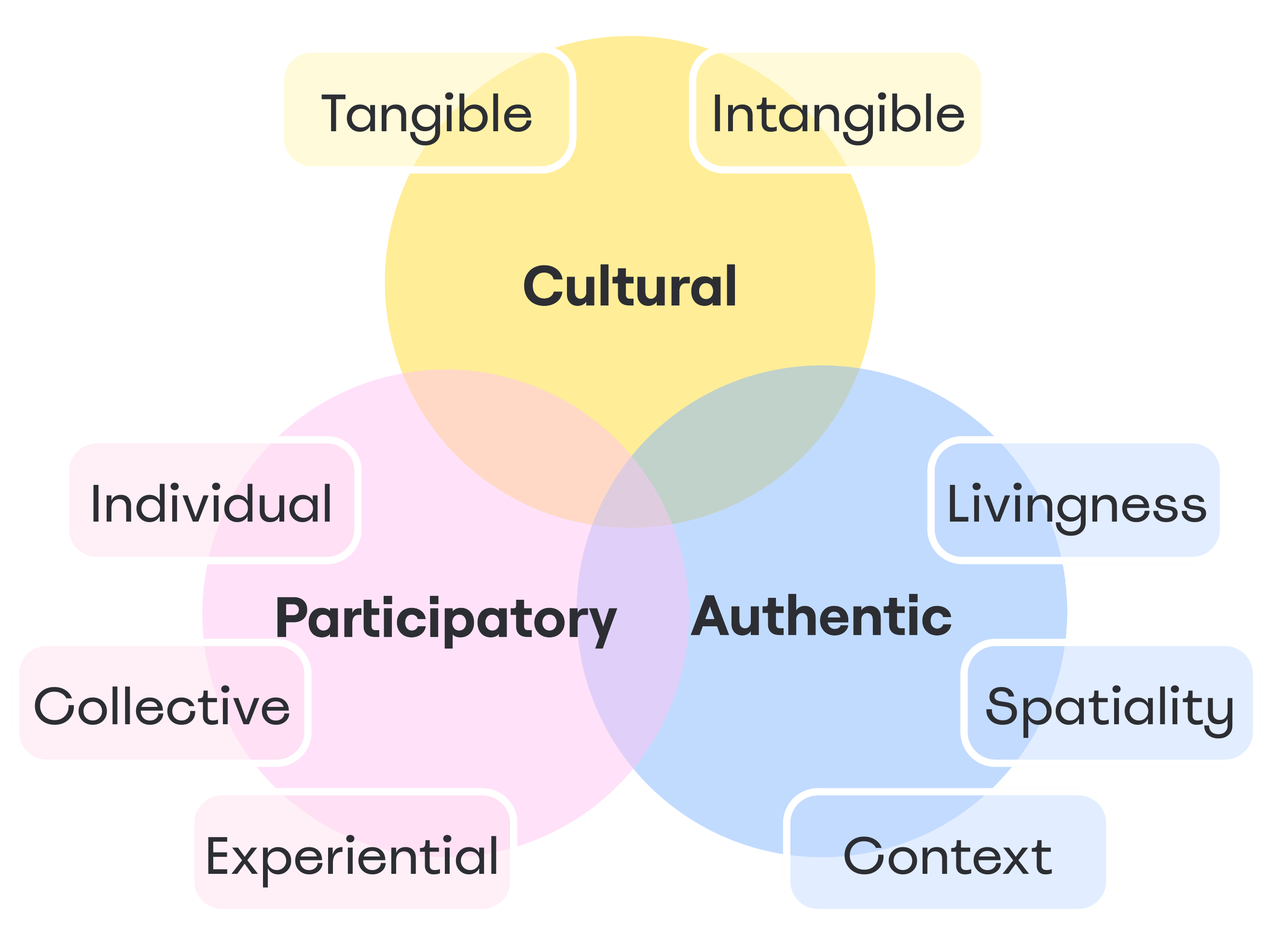}
  \Description{A Venn diagram with three elements. The first, “Cultural”, contains “Tangible” and “Intangible”. The second, “Participatory”, contains “Individual”, “Collective”, and “Experiential”. The third, “Authentic”, contains “Livingness”, “Spatiality”, and “Context”.}
  \caption{MRSLS as a heritage site interaction experience encompasses three aspects.}\label{fig:design-framework}
\end{figure}

Our findings reveal that MRSLS constitutes a unique medium for experiencing cultural heritage sites, distinguished by a triad of interconnected qualities: its capacity for cultural expression, its participatory nature, and its grounding in authenticity (\cref{fig:design-framework}).

First, it enables \textbf{cultural} engagement and showcase. On the one hand, \emph{tangible} cultural elements should be directly capturable by the camera, completing the depiction of the site's atmosphere.
Upon this living canvas, on the other hand, \emph{intangible} cultural elements are carefully curated and overlaid by designers---such as customs, poetry, and historical narratives. This fusion ensures that virtual elements are not detached artifacts but are harmoniously integrated into the site's real-world context, creating a coherent and meaningful cultural showcase.

Second, MRSLS deepens the \textbf{participatory} attributes inherent to live streaming.
\emph{Individually}, users gain a stronger sense of situatedness and telepresence even when they are not physically on site.
\emph{Collectively}, they engage in playful, lightweight social interactions that transform passive viewing into a shared, co-created experience.
\emph{Experientially}, they move beyond simply consuming cultural information to hands-on participation in collective cultural activities---something that is difficult to achieve in other non-VR remote applications.

Lastly, it spotlights the \textbf{authenticity} of the heritage site in multiple senses.
It preserves the \emph{livingness} of the stream by showing real-time events as they unfold.
It honors the \emph{spatiality} of the geographical environment by anchoring MR content to its physical features.
Lastly, it maintains the authenticity of \emph{context}, ensuring that both the view and the interactive cultural activities are true to the place itself.

With this simple framework as a generative guide, we encourage practitioners to leverage these three pillars when designing compelling and respectful MR functionalities for future MRSLS applications.

\subsection{Implications for Future Research and Design}

\subsubsection{Specializing MRSLS}
Our prototype was designed to be compatible with general-purpose live stream platforms to ensure familiarity and ease of use.
This approach, however, constrains interactions to standard live stream inputs like comments and gifts.
Future work could explore bespoke MRSLS platforms that offer richer and more nuanced user controls.
Such platforms could not only enable more flexible interactivity but also empower users to personalize their experience by filtering feature sets or adjusting visual styles, directly addressing the diverse preferences we observed (\cref{sec:eval:ar}).

\subsubsection{Articulating the Dynamic Present}
The current MRSLS prototype primarily interacts with the static elements of the scene, such as the landscape and architecture.
Yet, the site's ``dynamic present'', particularly the ambient activity of local inhabitants, is also a vital part of a lively place~\cite{halbwachsSpaceCollectiveMemory1950}.
Our design has yet to fully exploit this in the MR interactions.

Therefore, as a form of media founded upon a specific physical site,
future MRSLS can also strive to incorporate the dynamic present within the scene.
For instance, a future MRSLS at the West Lake can track the real boats on the water, allowing virtual lotuses to realistically collide into them.
\edit{Also, it can detect the scene's dynamic environmental context, enabling special interactions between virtual elements and lighting conditions, weathers, ambient sounds, etc.}
From an entertainment perspective, integrating such unpredictable, real-world events can create emergent and ever-fresh experiences, significantly boosting the ``revisitability'' of the MRSLS broadcast~\cite{adetunjiUnlockingLearningInvestigating2024}.

This concept could be similarly extended to create bidirectional interactions. By installing on-site IoT devices or public displays, local inhabitants could become aware of the MRSLS and the remote audience. This opens the possibility for them to interact with the virtual scene and its online participants via on-site installations, forging a direct connection between the physical and digital communities. Such a bidirectional link would add another layer of liveness. And it can also transform the experience for locals, offering \emph{them} a novel, live-streaming-like way to engage with a global audience beyond geographical boundaries.

\subsubsection{MRSLS and Digital Placemaking}
The articulation of the dynamic present leads us to a deeper reflection of how MRSLS participates in placemaking---the process by which communities imbue geographical ``spaces'' with shared meaning to create conceptual ``places''~\cite{aucoinAnthropologicalUnderstandingSpace2017}.
From this viewpoint, our work can be framed by the foundational HCI distinction between space and place~\cite{harrisonReplaceingSpaceRoles1996}. We argue that a plain SLS presents the heritage site as a mere geographical space---a passive, geometric backdrop for observation. In contrast, our MRSLS design provides the interactive and cultural layers necessary for a remote audience to transform that space into a shared place, imbuing it with collective meaning, activity, and social presence.
As SLS becomes an increasingly common window into places acorss the globe---for education, tourism, and homesickness relief~\cite{jarrattExplorationWebcamtravelConnecting2021,jarrattWebcamtravelConceptualFoundations2021,gonzalezKeepingStrongConnections2007}---we argue that this deliberate, culturally-aware approach is both meaningful and necessary.

In this context, several critical questions may arise regarding the potential radical future of MRSLS\@.
As local inhabitants become aware of their constant digital presence, might their behaviors and their relationship with the site change, ultimately reshaping the meaning of the place itself?
Conversely, as a global audience virtually inhabits this site through MRSLS, are they becoming new stakeholders in its placemaking process? The established tradition of placemaking being a local endeavor~\cite{almqvistDifferentTogetherDesign2023} is challenged when remote users can leave persistent digital traces.
Finally, with our call for a careful, culturally-sensitive design process, how should designers navigate the potential tensions that arise when their interpretation of a site's culture misaligns with that of the local community?

While some of these questions apply to plain SLS as well, our work on MRSLS brings them into sharper focus. Answering them will require longitudinal and sociological studies, which will, in turn, provide critical design considerations for the responsible development of this technology.

\subsection{Limitations and Future Work}
Despite the rich insights from our work, several limitations should be noted.
First, \edit{the scale of the studies was limited.
The formative interview only recruited six participants, which might impose limitations on the selection of MR elements.}
Also, all current studies focused on young adults aged 18--30, who are the primary demographic for (general) live streams in general~\cite{chenWhatDrivesLivestream2018}.
\edit{We used this age group during our studies to approximate that of scenic live streams, as the limited research on SLS hindered us from knowing its primary age group. Still, we acknowledge that this approximation may be potentially inaccurate.}
Concerning the cultural background, our current study exclusively focused on Chinese audiences due to the rich cultural context of the West Lake.
\edit{As an exploratory design research, we have gained substantial insights from this user group. With (MR)SLS gaining more attention in the HCI community,} future studies can engage larger-scale audience groups with more diverse age ranges and cultural backgrounds.
\edit{Longitudinal designs and cross-cultural comparisons are also suggested, as they can shed light on the appropriateness and generalizability of MRSLS.}

Second, our exploratory studies of this novel hybrid experience necessitated creating a customized survey.
While we ensured its comprehensiveness by synthesizing items from established questionnaires on place attachment~\cite{raymondMeasurementPlaceAttachment2010}, affective and cognitive ``immersion'' in entertaining systems~\cite{jennettMeasuringDefiningExperience2008}, gamified touristic information retrieval~\cite{sigalaApplicationImpactGamification2015}, and synchronous colocated XR games~\cite{bhattacharyyaBrickModelDesigning2019}, no single standardized instrument mentioned above perfectly matches the corresponding component in MRSLS.
For example, the concept of ``place connection'' in SLS explicitly lacks a formal measurement framework~\cite{gonzalezKeepingStrongConnections2007,jarrattExplorationWebcamtravelConnecting2021}; existing ``place attachment'' scales are more suited to places with prior personal history~\cite{lewickaPlaceAttachmentHow2011,raymondMeasurementPlaceAttachment2010,hidalgoPLACEATTACHMENTCONCEPTUAL2001}.
As (MR-)SLS gains more attention, we call for the development of a systematic evaluation framework tailored to these unique experiences.

\edit{Third, our current prototype is mainly implemented with generic Unity shaders and local assets.
As we have uncovered users' diverse preferences for aesthetic styles, future systems can incorporate more advanced visual techniques with precise aims.
For example, a realistic MRSLS can leverage high-fidelity assets and environment-aware lighting, while a cartoonish MRSLS can employ dramatic effects and AI-assisted art design.
Furthermore, concerning the requirement for immediate feedback and action confirmation (\cref{sec:features:fish}), future designs can introduce more feedback channels to reinforce agency and action-awareness. For example, outside the main live stream view, there could be a real-time updating queue, showing those commands and their issuers that have just been received but still pending network delay or animation throttling.}

Finally, our current research focuses exclusively on cultural heritage sites. Future studies can explore the generalizability of the MRSLS concept to other contexts, such as everyday urban districts, natural landscapes, or even commercial spaces like shopping malls.

\section{Conclusion}
This paper introduces and demonstrates Mixed Reality Scenic Live Streams (MRSLS), a new media form augmenting passive scenic live streams (SLS) with interactive, culturally relevant virtual content. We designed and evaluated an MRSLS prototype at a UNESCO-listed heritage site in China, transforming the viewing experience from passive observation into active participation. Our mixed-methods study confirms that, compared to plain SLS, MRSLS affords a stronger sense of connection to the remote site, fosters a more vibrant shared social experience, and enriches the scene with cultural and aesthetic value. By integrating the authenticity of real-time footage with the collective engagement of live streaming, MRSLS opens a compelling design space for remote cultural heritage interaction. We contend that this fusion of cultural, participatory, and authentic attributes offers a powerful new medium for cultural transmission in an increasingly digital world.


\begin{acks}
  This research is conducted in collaboration with \company{} We appreciate their engineering support and digital assets.

  This project is partially supported by the Hong Kong SAR Research Grants Council's Theme-based Research Grant Scheme (Project No. T45-407/19N).

  Icons used in \cref{fig:architecture} requires attributions.
  The film, comment, camera, broadcasting, and 3D modeling icons are designed by \href{https://www.freepik.com/}{Freepik}.
  The 3D scanner and paint icons are designed by Smashicons.
  The audience engagement icon is designed by Eucalyp.
  The gift icon is designed by Earthz Stocker.
  The umbrella, lotus, boat, fireworks, and koi icons are designed by \href{https://icons8.com/}{Icons8}.
\end{acks}

\bibliographystyle{ACM-Reference-Format}
\bibliography{acm}

@article{adetunjiUnlockingLearningInvestigating2024,
  title = {Unlocking {{Learning}}: {{Investigating}} the {{Replayability}} of {{Educational Games}}},
  shorttitle = {Unlocking {{Learning}}},
  author = {Adetunji, Rose Oluwaseun and {Ade-Ibijola}, Abejide},
  year = 2024,
  journal = {International Journal of Computer Games Technology},
  volume = {2024},
  number = {1},
  pages = {5876780},
  issn = {1687-7055},
  doi = {10.1155/2024/5876780},
  urldate = {2024-12-10},
  copyright = {Copyright \copyright{} 2024 Rose Oluwaseun Adetunji and Abejide Ade-Ibijola.},
  langid = {english}
}

@inproceedings{almqvistDifferentTogetherDesign2023,
  title = {Different {{Together}}: {{Design}} for {{Radical Placemaking}}},
  shorttitle = {Different {{Together}}},
  booktitle = {Proceedings of the 2023 {{CHI Conference}} on {{Human Factors}} in {{Computing Systems}}},
  author = {Almqvist, Andreas and Hedman, Anders and Clear, Adrian K and Comber, Rob},
  year = 2023,
  month = apr,
  series = {{{CHI}} '23},
  pages = {1--16},
  publisher = {Association for Computing Machinery},
  address = {New York, NY, USA},
  doi = {10.1145/3544548.3581080},
  urldate = {2024-05-29},
  isbn = {978-1-4503-9421-5}
}

@inproceedings{arendttorpGrabItYou2023,
  title = {Grab {{It}}, {{While You Can}}: {{A VR Gesture Evaluation}} of a {{Co-Designed Traditional Narrative}} by {{Indigenous People}}},
  shorttitle = {Grab {{It}}, {{While You Can}}},
  booktitle = {Proceedings of the 2023 {{CHI Conference}} on {{Human Factors}} in {{Computing Systems}}},
  author = {Arendttorp, Emilie Maria Nybo and {Winschiers-Theophilus}, Heike and Rodil, Kasper and Johansen, Freja B. K. and Rosengreen J{\o}rgensen, Mads and Kjeldsen, Thomas K. K. and Magot, Samkao},
  year = 2023,
  month = apr,
  pages = {1--13},
  publisher = {ACM},
  address = {Hamburg Germany},
  doi = {10.1145/3544548.3580894},
  urldate = {2025-11-10},
  isbn = {978-1-4503-9421-5},
  langid = {english}
}

@incollection{aucoinAnthropologicalUnderstandingSpace2017,
  title = {Toward an {{Anthropological Understanding}} of {{Space}} and {{Place}}},
  booktitle = {Place, {{Space}} and {{Hermeneutics}}},
  author = {Aucoin, Pauline McKenzie},
  editor = {Janz, Bruce B.},
  year = 2017,
  pages = {395--412},
  publisher = {Springer International Publishing},
  address = {Cham},
  doi = {10.1007/978-3-319-52214-2_28},
  urldate = {2024-12-10},
  isbn = {978-3-319-52214-2},
  langid = {english}
}

@inproceedings{baishyaYourEyesAnytime2017,
  title = {In {{Your Eyes}}: {{Anytime}}, {{Anywhere Video}} and {{Audio Streaming}} for {{Couples}}},
  shorttitle = {In {{Your Eyes}}},
  booktitle = {Proceedings of the 2017 {{ACM Conference}} on {{Computer Supported Cooperative Work}} and {{Social Computing}}},
  author = {Baishya, Uddipana and Neustaedter, Carman},
  year = 2017,
  month = feb,
  series = {{{CSCW}} '17},
  pages = {84--97},
  publisher = {Association for Computing Machinery},
  address = {New York, NY, USA},
  doi = {10.1145/2998181.2998200},
  urldate = {2023-01-04},
  isbn = {978-1-4503-4335-0}
}

@inproceedings{bhattacharyyaBrickModelDesigning2019,
  title = {Brick: {{Toward A Model}} for {{Designing Synchronous Colocated Augmented Reality Games}}},
  shorttitle = {Brick},
  booktitle = {Proceedings of the 2019 {{CHI Conference}} on {{Human Factors}} in {{Computing Systems}}},
  author = {Bhattacharyya, Po and Nath, Radha and Jo, Yein and Jadhav, Ketki and Hammer, Jessica},
  year = 2019,
  month = may,
  series = {{{CHI}} '19},
  pages = {1--9},
  publisher = {Association for Computing Machinery},
  address = {New York, NY, USA},
  doi = {10.1145/3290605.3300553},
  urldate = {2022-03-22},
  isbn = {978-1-4503-5970-2}
}

@article{braunUsingThematicAnalysis2006,
  title = {Using Thematic Analysis in Psychology},
  author = {Braun, Virginia and Clarke, Victoria},
  year = 2006,
  month = jan,
  journal = {Qualitative Research in Psychology},
  volume = {3},
  number = {2},
  pages = {77--101},
  publisher = {Routledge},
  issn = {1478-0887},
  doi = {10.1191/1478088706qp063oa},
  urldate = {2022-09-13}
}

@article{burkeyTotalRecallHow2019,
  title = {Total {{Recall}}: {{How Cultural Heritage Communities Use Digital Initiatives}} and {{Platforms}} for {{Collective Remembering}}},
  shorttitle = {Total {{Recall}}},
  author = {Burkey, Brant},
  year = 2019,
  month = nov,
  journal = {Journal of Creative Communications},
  volume = {14},
  number = {3},
  pages = {235--253},
  publisher = {SAGE Publications India},
  issn = {0973-2586},
  doi = {10.1177/0973258619868045},
  urldate = {2025-06-16},
  langid = {english}
}

@article{cejkaHybridAugmentedReality2020,
  title = {A Hybrid Augmented Reality Guide for Underwater Cultural Heritage Sites},
  author = {{\v C}ejka, Jan and Zs{\'i}ros, Attila and Liarokapis, Fotis},
  year = 2020,
  month = dec,
  journal = {Personal and Ubiquitous Computing},
  volume = {24},
  number = {6},
  pages = {815--828},
  issn = {1617-4917},
  doi = {10.1007/s00779-019-01354-6},
  urldate = {2022-06-17},
  langid = {english}
}

@article{chenWhatDrivesLivestream2018,
  title = {What Drives Live-Stream Usage Intention? {{The}} Perspectives of Flow, Entertainment, Social Interaction, and Endorsement},
  shorttitle = {What Drives Live-Stream Usage Intention?},
  author = {Chen, Chia-Chen and Lin, Yi-Chen},
  year = 2018,
  month = apr,
  journal = {Telematics and Informatics},
  volume = {35},
  number = {1},
  pages = {293--303},
  issn = {0736-5853},
  doi = {10.1016/j.tele.2017.12.003},
  urldate = {2022-07-08},
  langid = {english}
}

@article{cropleyPraiseConvergentThinking2006,
  title = {In {{Praise}} of {{Convergent Thinking}}},
  author = {Cropley, Arthur},
  year = 2006,
  month = jul,
  journal = {Creativity Research Journal},
  volume = {18},
  number = {3},
  pages = {391--404},
  issn = {1040-0419, 1532-6934},
  doi = {10.1207/s15326934crj1803_13},
  urldate = {2025-11-10},
  langid = {english}
}

@inproceedings{dengBlendedTourismExperiencescape2019,
  title = {Blended {{Tourism Experiencescape}}: {{A Conceptualisation}} of {{Live-Streaming Tourism}}},
  shorttitle = {Blended {{Tourism Experiencescape}}},
  booktitle = {Information and {{Communication Technologies}} in {{Tourism}} 2019},
  author = {Deng, Zhiming and Benckendorff, Pierre and Wang, Jie},
  editor = {Pesonen, Juho and Neidhardt, Julia},
  year = 2019,
  pages = {212--222},
  publisher = {Springer International Publishing},
  address = {Cham},
  doi = {10.1007/978-3-030-05940-8_17},
  isbn = {978-3-030-05940-8},
  langid = {english}
}

@article{gaoServiceDesignDestination2022,
  title = {Service Design for the Destination Tourism Service Ecosystem: A Review and Extension},
  shorttitle = {Service Design for the Destination Tourism Service Ecosystem},
  author = {Gao, Ying and Zhang, Qing and Xu, Xiaofeng and Jia, Fu and Lin, Zhibin},
  year = 2022,
  month = mar,
  journal = {Asia Pacific Journal of Tourism Research},
  volume = {27},
  number = {3},
  pages = {225--245},
  publisher = {Routledge},
  issn = {1094-1665},
  doi = {10.1080/10941665.2022.2046119},
  urldate = {2022-09-13}
}

@article{gattoImprovingAccessibilityCultural2025,
  title = {Improving Accessibility to Cultural Heritage: {{Integration}} of Extended Reality, Tactile Prints and User Experience Analysis for the Church of {{Madonna}} Dell'{{Itri}}},
  shorttitle = {Improving Accessibility to Cultural Heritage},
  author = {Gatto, Carola and Barba, Maria Cristina and Chiarello, Sofia and Corchia, Laura and Faggiano, Federica and Nuzzo, Benito Luigi and Riera Panaro, Ileana and Sumerano, Giada and De Luca, Valerio and De Giorgi, Manuela and De Paolis, Lucio Tommaso},
  year = 2025,
  month = may,
  journal = {J. Comput. Cult. Herit.},
  issn = {1556-4673},
  doi = {10.1145/3733154},
  urldate = {2025-06-25}
}

@article{genayBeingAvatarReal2022,
  title = {Being an {{Avatar}} "for {{Real}}": A {{Survey}} on {{Virtual Embodiment}} in {{Augmented Reality}}},
  shorttitle = {Being an {{Avatar}} "for {{Real}}"},
  author = {Genay, Adelaide Charlotte Sarah and Lecuyer, Anatole and Hachet, Martin},
  year = 2022,
  month = dec,
  journal = {IEEE Transactions on Visualization and Computer Graphics},
  volume = {28},
  number = {12},
  pages = {5071--5090},
  issn = {1941-0506},
  doi = {10.1109/TVCG.2021.3099290}
}

@article{geRelationshipSoundscapeChinese2012,
  title = {The {{Relationship Between Soundscape}} and {{Chinese Culture}} -- a {{Case Study}} of the {{West Lake Scenic Area}}},
  author = {Ge, J. and Guo, M. and Zhu, Y. H. and Jia, J. and Zhang, Y.},
  year = 2012,
  month = dec,
  journal = {Lowland Technology International},
  volume = {14},
  number = {2, Dec},
  pages = {23--37},
  issn = {2187-8870},
  urldate = {2022-12-24},
  copyright = {Copyright (c) 2019 Lowland Technology International},
  langid = {english}
}

@article{geSoundscapeWestLake2013,
  title = {Soundscape of the {{West Lake Scenic Area}} with Profound Cultural Background---a Case Study of {{Evening Bell Ringing}} in {{Jingci Temple}}, {{China}}},
  author = {Ge, Jian and Guo, Min and Yue, Miao},
  year = 2013,
  month = mar,
  journal = {Journal of Zhejiang University SCIENCE A},
  volume = {14},
  number = {3},
  pages = {219--229},
  issn = {1862-1775},
  doi = {10.1631/jzus.A1200159},
  urldate = {2022-09-02},
  langid = {english}
}

@article{gonzalezKeepingStrongConnections2007,
  title = {Keeping {{Strong Connections}} to the {{Homeland}} via {{Web-based Tools}}: {{The Case}} of {{Mexican Migrant Communities}} in the {{United States}}},
  shorttitle = {Keeping {{Strong Connections}} to the {{Homeland}} via {{Web-based Tools}}},
  author = {Gonz{\'a}lez, V{\'i}ctor M. and Castro, Luis A.},
  year = 2007,
  journal = {Journal of Community Informatics},
  volume = {3},
  number = {3},
  doi = {10.15353/joci.v3i3.2362}
}

@incollection{haimsonWhatMakesLive2017,
  title = {What {{Makes Live Events Engaging}} on {{Facebook Live}}, {{Periscope}}, and {{Snapchat}}},
  booktitle = {Proceedings of the 2017 {{CHI Conference}} on {{Human Factors}} in {{Computing Systems}}},
  author = {Haimson, Oliver L. and Tang, John C.},
  year = 2017,
  month = may,
  pages = {48--60},
  publisher = {Association for Computing Machinery},
  address = {New York, NY, USA},
  urldate = {2022-07-10},
  isbn = {978-1-4503-4655-9}
}

@incollection{halbwachsSpaceCollectiveMemory1950,
  title = {Space and the {{Collective Memory}}},
  booktitle = {The {{Collective Memory}}},
  author = {Halbwachs, Maurice},
  year = 1950,
  urldate = {2024-01-24}
}

@inproceedings{hamiltonStreamingTwitchFostering2014,
  title = {Streaming on Twitch: Fostering Participatory Communities of Play within Live Mixed Media},
  shorttitle = {Streaming on Twitch},
  booktitle = {Proceedings of the {{SIGCHI Conference}} on {{Human Factors}} in {{Computing Systems}}},
  author = {Hamilton, William A. and Garretson, Oliver and Kerne, Andruid},
  year = 2014,
  month = apr,
  series = {{{CHI}} '14},
  pages = {1315--1324},
  publisher = {Association for Computing Machinery},
  address = {New York, NY, USA},
  doi = {10.1145/2556288.2557048},
  urldate = {2023-01-04},
  isbn = {978-1-4503-2473-1}
}

@inproceedings{harrisonReplaceingSpaceRoles1996,
  title = {Re-Place-Ing Space: The Roles of Place and Space in Collaborative Systems},
  shorttitle = {Re-Place-Ing Space},
  booktitle = {Proceedings of the 1996 {{ACM}} Conference on {{Computer}} Supported Cooperative Work},
  author = {Harrison, Steve and Dourish, Paul},
  year = 1996,
  month = nov,
  series = {{{CSCW}} '96},
  pages = {67--76},
  publisher = {Association for Computing Machinery},
  address = {New York, NY, USA},
  doi = {10.1145/240080.240193},
  urldate = {2025-09-26},
  isbn = {978-0-89791-765-0}
}

@article{hidalgoPLACEATTACHMENTCONCEPTUAL2001,
  title = {{{PLACE ATTACHMENT}}: {{CONCEPTUAL AND EMPIRICAL QUESTIONS}}},
  shorttitle = {{{PLACE ATTACHMENT}}},
  author = {Hidalgo, M. CARMEN and Hern{\'a}ndez, {\relax BERNARDO}},
  year = 2001,
  month = sep,
  journal = {Journal of Environmental Psychology},
  volume = {21},
  number = {3},
  pages = {273--281},
  issn = {0272-4944},
  doi = {10.1006/jevp.2001.0221},
  urldate = {2022-06-12},
  langid = {english}
}

@inproceedings{huangBeThereBe2024,
  title = {Be {{There}}, {{Be Together}}, {{Be Streamed}}! {{AR Scenic Live-Streaming}} for an {{Interactive}} and {{Collective Experience}}},
  booktitle = {2024 {{IEEE International Symposium}} on {{Mixed}} and {{Augmented Reality Adjunct}} ({{ISMAR-Adjunct}})},
  author = {Huang, Zeyu and Xu, Zuyu and Zhang, Yuanhao and Liu, Chengzhong and Zhao, Yanwei and Shi, Chuhan and Zhao, Jason Chen and Ma, Xiaojuan},
  year = 2024,
  month = oct,
  pages = {453--456},
  publisher = {IEEE},
  issn = {2771-1110},
  doi = {10.1109/ISMAR-Adjunct64951.2024.00132},
  urldate = {2024-12-07}
}

@incollection{hutsonImmersiveTechnologies2024,
  title = {Immersive {{Technologies}}},
  booktitle = {Inclusive {{Smart Museums}}: {{Engaging Neurodiverse Audiences}} and {{Enhancing Cultural Heritage}}},
  author = {Hutson, James and Hutson, Piper},
  editor = {Hutson, James and Hutson, Piper},
  year = 2024,
  pages = {153--228},
  publisher = {Springer Nature Switzerland},
  address = {Cham},
  doi = {10.1007/978-3-031-43615-4_5},
  urldate = {2025-11-10},
  isbn = {978-3-031-43615-4},
  langid = {english}
}

@article{huWhyAudiencesChoose2017,
  title = {Why Do Audiences Choose to Keep Watching on Live Video Streaming Platforms? {{An}} Explanation of Dual Identification Framework},
  shorttitle = {Why Do Audiences Choose to Keep Watching on Live Video Streaming Platforms?},
  author = {Hu, Mu and Zhang, Mingli and Wang, Yu},
  year = 2017,
  month = oct,
  journal = {Computers in Human Behavior},
  volume = {75},
  pages = {594--606},
  issn = {0747-5632},
  doi = {10.1016/j.chb.2017.06.006},
  urldate = {2022-06-21},
  langid = {english}
}

@article{iacovielloHoloCitiesSharedReality2020,
  title = {{{HoloCities}}: {{A Shared Reality}} Application for {{Collaborative Tourism}}},
  shorttitle = {{{HoloCities}}},
  author = {Iacoviello, Roberto and Zappia, Davide},
  year = 2020,
  month = nov,
  journal = {IOP Conference Series: Materials Science and Engineering},
  volume = {949},
  number = {1},
  pages = {012036},
  publisher = {IOP Publishing},
  issn = {1757-899X},
  doi = {10.1088/1757-899X/949/1/012036},
  urldate = {2025-05-20},
  langid = {english}
}

@incollection{jacobCollaborativeAugmentedReality2021,
  title = {Collaborative {{Augmented Reality}} for {{Cultural Heritage}}, {{Tourist Sites}} and {{Museums}}: {{Sharing Visitors}}' {{Experiences}} and {{Interactions}}},
  shorttitle = {Collaborative {{Augmented Reality}} for {{Cultural Heritage}}, {{Tourist Sites}} and {{Museums}}},
  booktitle = {Augmented {{Reality}} in {{Tourism}}, {{Museums}} and {{Heritage}}: {{A New Technology}} to {{Inform}} and {{Entertain}}},
  author = {Jacob, Jo{\~a}o and N{\'o}brega, Rui},
  editor = {Geroimenko, Vladimir},
  year = 2021,
  pages = {27--47},
  publisher = {Springer International Publishing},
  address = {Cham},
  doi = {10.1007/978-3-030-70198-7_2},
  urldate = {2025-05-20},
  isbn = {978-3-030-70198-7},
  langid = {english}
}

@article{jarrattExplorationWebcamtravelConnecting2021,
  title = {An Exploration of Webcam-Travel: {{Connecting}} to Place and Nature through Webcams during the {{COVID-19}} Lockdown of 2020},
  shorttitle = {An Exploration of Webcam-Travel},
  author = {Jarratt, David},
  year = 2021,
  month = apr,
  journal = {Tourism and Hospitality Research},
  volume = {21},
  number = {2},
  pages = {156--168},
  publisher = {SAGE Publications},
  issn = {1467-3584},
  doi = {10.1177/1467358420963370},
  urldate = {2022-06-22},
  langid = {english}
}

@article{jarrattWebcamtravelConceptualFoundations2021,
  title = {Webcam-Travel: {{Conceptual}} Foundations},
  shorttitle = {Webcam-Travel},
  author = {Jarratt, David},
  year = 2021,
  month = nov,
  journal = {Annals of Tourism Research},
  volume = {91},
  pages = {103088},
  issn = {0160-7383},
  doi = {10.1016/j.annals.2020.103088},
  urldate = {2022-09-06},
  langid = {english}
}

@article{jennettMeasuringDefiningExperience2008,
  title = {Measuring and Defining the Experience of Immersion in Games},
  author = {Jennett, Charlene and Cox, Anna L. and Cairns, Paul and Dhoparee, Samira and Epps, Andrew and Tijs, Tim and Walton, Alison},
  year = 2008,
  month = sep,
  journal = {International Journal of Human-Computer Studies},
  volume = {66},
  number = {9},
  pages = {641--661},
  issn = {1071-5819},
  doi = {10.1016/j.ijhcs.2008.04.004},
  urldate = {2022-09-08},
  langid = {english}
}

@article{jinExperientialAuthenticityHeritage2020,
  title = {Experiential Authenticity in Heritage Museums},
  author = {Jin, Liuhe and Xiao, Honggen and Shen, Haili},
  year = 2020,
  month = dec,
  journal = {Journal of Destination Marketing \& Management},
  volume = {18},
  pages = {100493},
  issn = {2212-571X},
  doi = {10.1016/j.jdmm.2020.100493},
  urldate = {2022-09-13},
  langid = {english}
}

@article{kilteniSenseEmbodimentVirtual2012,
  title = {The {{Sense}} of {{Embodiment}} in {{Virtual Reality}}},
  author = {Kilteni, Konstantina and Groten, Raphaela and Slater, Mel},
  year = 2012,
  journal = {Presence},
  volume = {21},
  number = {4},
  pages = {373--387},
  issn = {1054-7460},
  doi = {10.1162/PRES_a_00124}
}

@inproceedings{kimWhoWillSubscribe2019,
  title = {Who Will {{Subscribe}} to {{My Streaming Channel}}? {{The Case}} of {{Twitch}}},
  shorttitle = {Who Will {{Subscribe}} to {{My Streaming Channel}}?},
  booktitle = {Conference {{Companion Publication}} of the 2019 on {{Computer Supported Cooperative Work}} and {{Social Computing}}},
  author = {Kim, Jina and Bae, Kunwoo and Park, Eunil and {del Pobil}, Angel P.},
  year = 2019,
  month = nov,
  series = {{{CSCW}} '19},
  pages = {247--251},
  publisher = {Association for Computing Machinery},
  address = {New York, NY, USA},
  doi = {10.1145/3311957.3359470},
  urldate = {2023-01-04},
  isbn = {978-1-4503-6692-2}
}

@inproceedings{kotutTrailHeritageSafeguarding2021,
  title = {Trail as {{Heritage}}: {{Safeguarding Location-Specific}} and {{Transient Indigenous Knowledge}}},
  shorttitle = {Trail as {{Heritage}}},
  booktitle = {3rd {{African Human-Computer Interaction Conference}}: {{Inclusiveness}} and {{Empowerment}}},
  author = {Kotut, Lindah and McCrickard, Scott D},
  year = 2021,
  month = jul,
  series = {{{AfriCHI}} 2021},
  pages = {94--102},
  publisher = {Association for Computing Machinery},
  address = {New York, NY, USA},
  doi = {10.1145/3448696.3448702},
  urldate = {2022-12-09},
  isbn = {978-1-4503-8869-6}
}

@article{koukopoulosEvaluatingUsabilityPersonal2019,
  title = {Evaluating the {{Usability}} and the {{Personal}} and {{Social Acceptance}} of a {{Participatory Digital Platform}} for {{Cultural Heritage}}},
  author = {Koukopoulos, Zois and Koukopoulos, Dimitrios},
  year = 2019,
  month = mar,
  journal = {Heritage},
  volume = {2},
  number = {1},
  pages = {1--26},
  publisher = {Multidisciplinary Digital Publishing Institute},
  issn = {2571-9408},
  doi = {10.3390/heritage2010001},
  urldate = {2025-06-16},
  copyright = {http://creativecommons.org/licenses/by/3.0/},
  langid = {english}
}

@article{langConstructionIntangibleCultural2019,
  title = {Construction of {{Intangible Cultural Heritage Spot Based}} on {{AR Technology}}---{{Taking}} the {{Intangible Cultural Heritage}} of the {{Li Nationality}} in the {{Areca Valley}} as {{An Example}}},
  author = {Lang, Yue and Deng, Xi and Zhang, Kun and Wang, Yiwen},
  year = 2019,
  journal = {IOP Conference Series: Earth and Environmental Science},
  volume = {234},
  pages = {012119},
  publisher = {IOP Publishing},
  issn = {1755-1315},
  doi = {10.1088/1755-1315/234/1/012119},
  urldate = {2022-07-08},
  langid = {english}
}

@inproceedings{leeUnderstandingHowDigital2019,
  title = {Understanding {{How Digital Gifting Influences Social Interaction}} on {{Live Streams}}},
  booktitle = {Proceedings of the 21st {{International Conference}} on {{Human-Computer Interaction}} with {{Mobile Devices}} and {{Services}}},
  author = {Lee, Yi-Chieh and Yen, Chi-Hsien and Wang, Dennis and Fu, Wai-Tat},
  year = 2019,
  month = oct,
  series = {{{MobileHCI}} '19},
  pages = {1--10},
  publisher = {Association for Computing Machinery},
  address = {New York, NY, USA},
  doi = {10.1145/3338286.3340144},
  urldate = {2022-09-13},
  isbn = {978-1-4503-6825-4}
}

@article{lewickaPlaceAttachmentHow2011,
  title = {Place Attachment: {{How}} Far Have We Come in the Last 40 Years?},
  shorttitle = {Place Attachment},
  author = {Lewicka, Maria},
  year = 2011,
  month = sep,
  journal = {Journal of Environmental Psychology},
  volume = {31},
  number = {3},
  pages = {207--230},
  issn = {0272-4944},
  doi = {10.1016/j.jenvp.2010.10.001},
  urldate = {2022-06-11},
  langid = {english}
}

@article{liExaminingGiftingBehavior2021,
  title = {Examining Gifting Behavior on Live Streaming Platforms: {{An}} Identity-Based Motivation Model},
  shorttitle = {Examining Gifting Behavior on Live Streaming Platforms},
  author = {Li, Ran and Lu, Yaobin and Ma, Jifeng and Wang, Weiquan},
  year = 2021,
  month = sep,
  journal = {Information \& Management},
  volume = {58},
  number = {6},
  pages = {103406},
  issn = {03787206},
  doi = {10.1016/j.im.2020.103406},
  urldate = {2022-06-09},
  langid = {english}
}

@article{linLiveStreamingTourism2022,
  title = {Live Streaming in Tourism and Hospitality: A Literature Review},
  shorttitle = {Live Streaming in Tourism and Hospitality},
  author = {Lin, Katsy and Fong, Lawrence Hoc Nang and Law, Rob},
  year = 2022,
  month = mar,
  journal = {Asia Pacific Journal of Tourism Research},
  volume = {27},
  number = {3},
  pages = {290--304},
  publisher = {Routledge},
  issn = {1094-1665},
  doi = {10.1080/10941665.2022.2061365},
  urldate = {2022-06-24}
}

@article{liSystematicReviewLiterature2020,
  title = {A {{Systematic Review}} of {{Literature}} on {{User Behavior}} in {{Video Game Live Streaming}}},
  author = {Li, Yi and Wang, Chongli and Liu, Jing},
  year = 2020,
  month = jan,
  journal = {International Journal of Environmental Research and Public Health},
  volume = {17},
  number = {9},
  pages = {3328},
  publisher = {Multidisciplinary Digital Publishing Institute},
  issn = {1660-4601},
  doi = {10.3390/ijerph17093328},
  urldate = {2022-06-10},
  copyright = {http://creativecommons.org/licenses/by/3.0/},
  langid = {english}
}

@article{liuAugmentedRealityassistedIntelligent2017,
  title = {Augmented {{Reality-assisted Intelligent Window}} for {{Cyber-Physical Machine Tools}}},
  author = {Liu, Chao and Cao, Sheng and Tse, Wayne and Xu, Xun},
  year = 2017,
  month = jul,
  journal = {Journal of Manufacturing Systems},
  series = {Special {{Issue}} on {{Latest}} Advancements in Manufacturing Systems at {{NAMRC}} 45},
  volume = {44},
  pages = {280--286},
  issn = {0278-6125},
  doi = {10.1016/j.jmsy.2017.04.008},
  urldate = {2024-06-27}
}

@inproceedings{luFeelItMy2019,
  title = {"{{I}} Feel It Is My Responsibility to Stream": {{Streaming}} and {{Engaging}} with {{Intangible Cultural Heritage}} through {{Livestreaming}}},
  shorttitle = {"{{I}} Feel It Is My Responsibility to Stream"},
  booktitle = {Proceedings of the 2019 {{CHI Conference}} on {{Human Factors}} in {{Computing Systems}}},
  author = {Lu, Zhicong and Annett, Michelle and Fan, Mingming and Wigdor, Daniel},
  year = 2019,
  month = may,
  series = {{{CHI}} '19},
  pages = {1--14},
  publisher = {Association for Computing Machinery},
  address = {New York, NY, USA},
  doi = {10.1145/3290605.3300459},
  urldate = {2022-12-09},
  isbn = {978-1-4503-5970-2}
}

@inproceedings{luRevivingEustonArch2023,
  title = {Reviving the {{Euston Arch}}: {{A Mixed Reality Approach}} to {{Cultural Heritage Tours}}},
  shorttitle = {Reviving the {{Euston Arch}}},
  booktitle = {2023 {{IEEE International Symposium}} on {{Mixed}} and {{Augmented Reality Adjunct}} ({{ISMAR-Adjunct}})},
  author = {Lu, Ziwen and Zhang, Jingyi and Shapiro, Kalila and Numan, Nels and Julier, Simon and Steed, Anthony},
  year = 2023,
  month = oct,
  pages = {821--826},
  issn = {2771-1110},
  doi = {10.1109/ISMAR-Adjunct60411.2023.00181},
  urldate = {2025-05-20}
}

@article{luVicariouslyExperiencingIt2019,
  title = {Vicariously {{Experiencing}} It All {{Without Going Outside}}: {{A Study}} of {{Outdoor Livestreaming}} in {{China}}},
  shorttitle = {Vicariously {{Experiencing}} It All {{Without Going Outside}}},
  author = {Lu, Zhicong and Annett, Michelle and Wigdor, Daniel},
  year = 2019,
  month = nov,
  journal = {Proceedings of the ACM on Human-Computer Interaction},
  volume = {3},
  number = {CSCW},
  pages = {25:1--25:28},
  doi = {10.1145/3359127},
  urldate = {2022-12-09}
}

@inproceedings{luYouWatchYou2018,
  title = {You {{Watch}}, {{You Give}}, and {{You Engage}}: {{A Study}} of {{Live Streaming Practices}} in {{China}}},
  shorttitle = {You {{Watch}}, {{You Give}}, and {{You Engage}}},
  booktitle = {Proceedings of the 2018 {{CHI Conference}} on {{Human Factors}} in {{Computing Systems}}},
  author = {Lu, Zhicong and Xia, Haijun and Heo, Seongkook and Wigdor, Daniel},
  year = 2018,
  month = apr,
  series = {{{CHI}} '18},
  pages = {1--13},
  publisher = {Association for Computing Machinery},
  address = {New York, NY, USA},
  doi = {10.1145/3173574.3174040},
  urldate = {2022-06-09},
  isbn = {978-1-4503-5620-6}
}

@phdthesis{marquezseguraBodyGamesDesigning2013,
  title = {Body {{Games}} : {{Designing}} for Movement-Based Play in Co-Located Social Settings.},
  shorttitle = {Body {{Games}}},
  author = {M{\'a}rquez Segura, Elena},
  year = 2013,
  month = dec
}

@misc{OilPaperUmbrellaMaking,
  title = {Umbrella {{Making Technique}} ({{Oil Paper Umbrella Making Technique}}) [in {{Chinese}}]},
  author = {{Chinese National Academy of Arts}},
  year = 2008,
  journal = {China Intangible Cultural Heritage Website [in Chinese]},
  urldate = {2025-06-25},
  howpublished = {https://www.ihchina.cn/project\_details/14579.html}
}

@article{qiuCanLiveStreaming2021,
  title = {Can {{Live Streaming Save}} the {{Tourism Industry}} from a {{Pandemic}}? {{A Study}} of {{Social Media}}},
  shorttitle = {Can {{Live Streaming Save}} the {{Tourism Industry}} from a {{Pandemic}}?},
  author = {Qiu, Qihang and Zuo, Yifan and Zhang, Mu},
  year = 2021,
  month = sep,
  journal = {ISPRS International Journal of Geo-Information},
  volume = {10},
  number = {9},
  pages = {595},
  publisher = {Multidisciplinary Digital Publishing Institute},
  issn = {2220-9964},
  doi = {10.3390/ijgi10090595},
  urldate = {2022-06-23},
  copyright = {http://creativecommons.org/licenses/by/3.0/},
  langid = {english}
}

@article{raymondMeasurementPlaceAttachment2010,
  title = {The {{Measurement}} of {{Place Attachment}}: {{Personal}}, {{Community}}, and {{Environmental Connections}}},
  shorttitle = {The {{Measurement}} of {{Place Attachment}}},
  author = {Raymond, Christopher and Brown, Gregory and Weber, Delene},
  year = 2010,
  month = dec,
  journal = {Journal of Environmental Psychology},
  volume = {30},
  number = {4},
  pages = {422--434},
  issn = {0272-4944},
  doi = {10.1016/j.jenvp.2010.08.002}
}

@inproceedings{renRedesignTraditionalChinese2013,
  title = {Redesign of Traditional {{Chinese}} Umbrella},
  booktitle = {Learn {{X Design Conference Series}}},
  author = {Ren, Lisha and Gao, Fengyu},
  year = 2013,
  month = sep,
  pages = {795--809},
  publisher = {ABM-media},
  address = {Oslo, Norway},
  doi = {10.21606/learnxdesign.2013.122}
}

@inproceedings{robinsonChatHasNo2022,
  title = {''{{Chat Has No Chill}}'': {{A Novel Physiological Interaction For Engaging Live Streaming Audiences}}},
  shorttitle = {''{{Chat Has No Chill}}''},
  booktitle = {{{CHI Conference}} on {{Human Factors}} in {{Computing Systems}}},
  author = {Robinson, Raquel Breejon and Rheeder, Ricardo and Klarkowski, Madison and Mandryk, Regan L},
  year = 2022,
  month = apr,
  pages = {1--18},
  publisher = {ACM},
  address = {New Orleans LA USA},
  doi = {10.1145/3491102.3501934},
  urldate = {2022-09-02},
  isbn = {978-1-4503-9157-3},
  langid = {english}
}

@article{romanVirtualSpaceTourism2022,
  title = {Virtual and {{Space Tourism}} as {{New Trends}} in {{Travelling}} at the {{Time}} of the {{COVID-19 Pandemic}}},
  author = {Roman, Micha{\l} and Kosi{\'n}ski, Robert and Bhatta, Kumar and Niedzi{\'o}{\l}ka, Arkadiusz and Krasnod{\k e}bski, Andrzej},
  year = 2022,
  month = jan,
  journal = {Sustainability},
  volume = {14},
  number = {2},
  pages = {628},
  publisher = {Multidisciplinary Digital Publishing Institute},
  issn = {2071-1050},
  doi = {10.3390/su14020628},
  urldate = {2022-06-13},
  copyright = {http://creativecommons.org/licenses/by/3.0/},
  langid = {english}
}

@inproceedings{scharffhausenWhatIfWe2024,
  title = {What If? We Could Virtually Restore and Visit a Pre-War Church to the Highest Fidelity Possible with No {{3D}} Scanning Available?},
  shorttitle = {What If?},
  booktitle = {Proceedings of the 26th {{Symposium}} on {{Virtual}} and {{Augmented Reality}}},
  author = {Scharffhausen, Jean-Baptiste and Teixeira, Jo{\~a}o Marcelo X. N.},
  year = 2024,
  month = sep,
  series = {{{SVR}} '24},
  pages = {329--333},
  publisher = {Association for Computing Machinery},
  address = {New York, NY, USA},
  doi = {10.1145/3691573.3691605},
  urldate = {2025-06-24},
  isbn = {979-8-4007-0979-1}
}

@article{shiCharacteristicsDevelopmentSignificance2015,
  title = {On the {{Characteristics}} and the {{Development Significance}} of {{Hangzhou Lotus Culture}}},
  author = {Shi, Hongbin},
  year = 2015,
  journal = {SHS Web of Conferences},
  volume = {14},
  pages = {02002},
  publisher = {EDP Sciences},
  issn = {2261-2424},
  doi = {10.1051/shsconf/20151402002},
  urldate = {2022-09-02},
  copyright = {\copyright{} Owned by the authors, published by EDP Sciences, 2015},
  langid = {english}
}

@article{siddiquiVirtualTourismDigital2022,
  title = {Virtual {{Tourism}} and {{Digital Heritage}}: {{An Analysis}} of {{VR}}/{{AR Technologies}} and {{Applications}}},
  shorttitle = {Virtual {{Tourism}} and {{Digital Heritage}}},
  author = {Siddiqui, Muhammad Shoaib and Syed, Toqeer Ali and Nadeem, Adnan and Nawaz, Waqas and Alkhodre, Ahmad},
  year = {2022/30/31},
  journal = {International Journal of Advanced Computer Science and Applications (IJACSA)},
  volume = {13},
  number = {7},
  publisher = {{The Science and Information (SAI) Organization Limited}},
  issn = {2156-5570},
  doi = {10.14569/IJACSA.2022.0130739},
  urldate = {2025-06-16},
  langid = {english}
}

@article{sigalaApplicationImpactGamification2015,
  title = {The Application and Impact of Gamification Funware on Trip Planning and Experiences: The Case of {{TripAdvisor}}'s Funware},
  shorttitle = {The Application and Impact of Gamification Funware on Trip Planning and Experiences},
  author = {Sigala, Marianna},
  year = 2015,
  month = sep,
  journal = {Electronic Markets},
  volume = {25},
  number = {3},
  pages = {189--209},
  issn = {1422-8890},
  doi = {10.1007/s12525-014-0179-1},
  urldate = {2022-09-14},
  langid = {english}
}

@article{sjoblomWhyPeopleWatch2017,
  title = {Why Do People Watch Others Play Video Games? {{An}} Empirical Study on the Motivations of {{Twitch}} Users},
  shorttitle = {Why Do People Watch Others Play Video Games?},
  author = {Sj{\"o}blom, Max and Hamari, Juho},
  year = 2017,
  month = oct,
  journal = {Computers in Human Behavior},
  volume = {75},
  pages = {985--996},
  issn = {0747-5632},
  doi = {10.1016/j.chb.2016.10.019},
  urldate = {2022-06-21},
  langid = {english}
}

@inproceedings{spierlingExperiencingPresenceHistorical2017,
  title = {Experiencing the {{Presence}} of {{Historical Stories}} with {{Location-Based Augmented Reality}}},
  booktitle = {Interactive {{Storytelling}}},
  author = {Spierling, Ulrike and Winzer, Peter and Massarczyk, Erik},
  editor = {Nunes, Nuno and Oakley, Ian and Nisi, Valentina},
  year = 2017,
  series = {Lecture {{Notes}} in {{Computer Science}}},
  pages = {49--62},
  publisher = {Springer International Publishing},
  address = {Cham},
  doi = {10.1007/978-3-319-71027-3_5},
  isbn = {978-3-319-71027-3},
  langid = {english}
}

@article{sylaiouVirtualHumansMuseums2022,
  title = {Virtual {{Humans}} in {{Museums}} and {{Cultural Heritage Sites}}},
  author = {Sylaiou, Stella and Fidas, Christos},
  year = 2022,
  month = jan,
  journal = {Applied Sciences},
  volume = {12},
  number = {19},
  pages = {9913},
  publisher = {Multidisciplinary Digital Publishing Institute},
  issn = {2076-3417},
  doi = {10.3390/app12199913},
  urldate = {2025-11-10},
  copyright = {http://creativecommons.org/licenses/by/3.0/},
  langid = {english}
}

@inproceedings{szentandrasiPOSTERINCASTInteractive2015,
  title = {[{{POSTER}}] {{INCAST}}: {{Interactive Camera Streams}} for {{Surveillance Cams AR}}},
  shorttitle = {[{{POSTER}}] {{INCAST}}},
  booktitle = {2015 {{IEEE International Symposium}} on {{Mixed}} and {{Augmented Reality}}},
  author = {Szentandr{\'a}si, I. and Zachari{\'a}, M. and Kajan, R. and Tinka, J. and Dubsk{\'a}, M. and Sochor, J. and Herout, A.},
  year = 2015,
  month = sep,
  series = {{{ISMAR}} '15},
  pages = {80--83},
  publisher = {IEEE Computer Society},
  address = {USA},
  doi = {10.1109/ISMAR.2015.26},
  isbn = {978-1-4673-7660-0}
}

@misc{timothys.y.lammuseumofanthropologyChineseNewYear2021,
  title = {Chinese {{New Year Traditions}}},
  author = {{Timothy S. Y. Lam Museum of Anthropology}},
  year = 2021,
  journal = {Wake Forest University},
  urldate = {2024-05-03},
  langid = {english}
}

@article{tongApplyingCinematicVirtual2024,
  title = {Applying {{Cinematic Virtual Reality}} with {{Adaptability}} to {{Indigenous Storytelling}}},
  author = {Tong, Lingwei and Lindeman, Robert W. and Lukosch, Heide and Clifford, Rory and Regenbrecht, Holger},
  year = 2024,
  month = mar,
  journal = {J. Comput. Cult. Herit.},
  volume = {17},
  number = {2},
  pages = {28:1--28:25},
  issn = {1556-4673},
  doi = {10.1145/3647996},
  urldate = {2025-06-25}
}

@incollection{tungAugmentedRealityMobile2015,
  title = {Augmented {{Reality}} for {{Mobile Service}} of {{Film-Induced Tourism App}}},
  booktitle = {Mobile {{Services}} for {{Toy Computing}}},
  author = {Tung, Wei-Feng},
  editor = {Hung, Patrick C. K.},
  year = 2015,
  series = {International {{Series}} on {{Computer Entertainment}} and {{Media Technology}}},
  pages = {129--140},
  publisher = {Springer International Publishing},
  address = {Cham},
  doi = {10.1007/978-3-319-21323-1_7},
  urldate = {2022-07-08},
  isbn = {978-3-319-21323-1},
  langid = {english}
}

@article{vertUserEvaluationMultiPlatform2021,
  title = {User {{Evaluation}} of a {{Multi-Platform Digital Storytelling Concept}} for {{Cultural Heritage}}},
  author = {Vert, Silviu and Andone, Diana and Ternauciuc, Andrei and Mihaescu, Vlad and Rotaru, Oana and Mocofan, Muguras and Orhei, Ciprian and Vasiu, Radu},
  year = 2021,
  month = jan,
  journal = {Mathematics},
  volume = {9},
  number = {21},
  pages = {2678},
  publisher = {Multidisciplinary Digital Publishing Institute},
  issn = {2227-7390},
  doi = {10.3390/math9212678},
  urldate = {2025-06-16},
  copyright = {http://creativecommons.org/licenses/by/3.0/},
  langid = {english}
}

@inproceedings{wangIntangibleCulturalHeritage2018,
  title = {The {{Intangible Cultural Heritage Show Mode Based}} on {{AR Technology}} in {{Museums}} - {{Take}} the {{Li Nationality Non-Material Cultural Heritage}} as an {{Example}}},
  booktitle = {2018 {{IEEE}} 3rd {{International Conference}} on {{Image}}, {{Vision}} and {{Computing}} ({{ICIVC}})},
  author = {Wang, Yiwen and Deng, Xi and Zhang, Kun and Lang, Yue},
  year = 2018,
  month = jun,
  pages = {936--940},
  publisher = {IEEE},
  address = {Chongqing},
  doi = {10.1109/ICIVC.2018.8492843},
  urldate = {2022-09-13},
  isbn = {978-1-5386-4991-6}
}

@inproceedings{webbDistributedLivenessUnderstanding2016,
  title = {Distributed {{Liveness}}: {{Understanding How New Technologies Transform Performance Experiences}}},
  shorttitle = {Distributed {{Liveness}}},
  booktitle = {Proceedings of the 19th {{ACM Conference}} on {{Computer-Supported Cooperative Work}} \& {{Social Computing}}},
  author = {Webb, Andrew M. and Wang, Chen and Kerne, Andruid and Cesar, Pablo},
  year = 2016,
  month = feb,
  series = {{{CSCW}} '16},
  pages = {432--437},
  publisher = {Association for Computing Machinery},
  address = {New York, NY, USA},
  doi = {10.1145/2818048.2819974},
  urldate = {2023-01-04},
  isbn = {978-1-4503-3592-8}
}

@misc{worldheritagecentreStateConservationProperties2019,
  title = {State of Conservation of Properties Inscribed on the {{World Heritage List}}},
  author = {{World Heritage Centre}},
  editor = {{UNESCO}},
  year = 2019,
  month = jun,
  publisher = {UNESCO},
  urldate = {2022-08-29},
  copyright = {UNESCO},
  langid = {english}
}

@article{wulfWatchingPlayersExploration2020,
  title = {Watching {{Players}}: {{An Exploration}} of {{Media Enjoyment}} on {{{\emph{Twitch}}}}},
  shorttitle = {Watching {{Players}}},
  author = {Wulf, Tim and Schneider, Frank M. and Beckert, Stefan},
  year = 2020,
  month = may,
  journal = {Games and Culture},
  volume = {15},
  number = {3},
  pages = {328--346},
  issn = {1555-4120, 1555-4139},
  doi = {10.1177/1555412018788161},
  urldate = {2022-06-21},
  langid = {english}
}

@article{xuHowImagesInspire2018,
  title = {How {{Images Inspire Poems}}: {{Generating Classical Chinese Poetry}} from {{Images}} with {{Memory Networks}}},
  shorttitle = {How {{Images Inspire Poems}}},
  author = {Xu, Linli and Jiang, Liang and Qin, Chuan and Wang, Zhe and Du, Dongfang},
  year = 2018,
  month = apr,
  journal = {Proceedings of the AAAI Conference on Artificial Intelligence},
  volume = {32},
  number = {1},
  pages = {5618--5625},
  issn = {2374-3468},
  doi = {10.1609/aaai.v32i1.12001},
  urldate = {2022-09-13},
  chapter = {Main Track: NLP and Machine Learning},
  copyright = {Copyright (c)},
  langid = {english}
}

@article{xuSignificanceWestLake2017,
  title = {The Significance of the {{West Lake}} Pattern and Its Heuristic Implications for Creating {{China}}'s Heritage Tourism Economics},
  author = {Xu, Songling and Liu, Yu and Qian, Yihong and Wang, Qiuju},
  year = 2017,
  month = feb,
  journal = {Tourism Management},
  volume = {58},
  pages = {286--292},
  issn = {0261-5177},
  doi = {10.1016/j.tourman.2016.03.013},
  urldate = {2022-09-02},
  langid = {english}
}

@article{zhangCanLiveStreaming2021,
  title = {Can ``{{Live Streaming}}'' {{Really Drive Visitors}} to the {{Destination}}? {{From}} the {{Aspect}} of ``{{Social Presence}}''},
  shorttitle = {Can ``{{Live Streaming}}'' {{Really Drive Visitors}} to the {{Destination}}?},
  author = {Zhang, Wenkun and Wang, Yanan and Zhang, Tao},
  year = 2021,
  month = jan,
  journal = {SAGE Open},
  volume = {11},
  number = {1},
  pages = {21582440211006691},
  publisher = {SAGE Publications},
  issn = {2158-2440},
  doi = {10.1177/21582440211006691},
  urldate = {2022-06-23},
  langid = {english}
}

@article{zhangCulturalLandscapeMeanings2019,
  title = {Cultural Landscape Meanings. {{The}} Case of {{West Lake}}, {{Hangzhou}}, {{China}}},
  author = {Zhang, Rouran and Taylor, Ken},
  year = 2019,
  month = mar,
  journal = {Landscape Research},
  volume = {45},
  pages = {1--15},
  doi = {10.1080/01426397.2019.1589438}
}

@article{zhouMagicDanmakuSocial2019,
  title = {The Magic of Danmaku: {{A}} Social Interaction Perspective of Gift Sending on Live Streaming Platforms},
  shorttitle = {The Magic of Danmaku},
  author = {Zhou, Jilei and Zhou, Jing and Ding, Ying and Wang, Hansheng},
  year = 2019,
  month = mar,
  journal = {Electronic Commerce Research and Applications},
  volume = {34},
  pages = {100815},
  issn = {1567-4223},
  doi = {10.1016/j.elerap.2018.11.002},
  urldate = {2022-06-09},
  langid = {english}
}

\end{document}